\documentclass[%
 reprint,
 amsmath,amssymb,
 aps,
]{revtex4-1}

\usepackage{subfigure}
\usepackage{footmisc}
\usepackage{graphicx}
\usepackage{dcolumn}
\usepackage{bm}

\newcommand{\opr}[1]{\ensuremath{\mathbf{\mathsf{#1}}}}

\newcommand{\abs}[1]{\ensuremath{\left|#1\right|}}

\newcommand{\kernel}[1]{\ensuremath{\left<q\left|#1\right|q'\right>}}

\newcommand{\dirac}[2]{\ensuremath{\left<#1\left|\right.#2\right>}}

\newcommand{\expec}[3]{\ensuremath{\left.\left<#1\left|#2\right|#3\right>\right.}}

\begin{document}

\preprint{APS/123-QED}

\title{Quantum free fall motion and quantum violation of weak equivalence principle}

\author{Philip Caesar M. Flores}
 \altaffiliation[]{pflores@nip.upd.edu.ph}
\author{Eric A. Galapon}%
 \email{eagalapon@up.edu.ph}
\affiliation{%
Theoretical Physics Group, National Institute of Physics, University of the Philippines, Diliman Quezon City, 1101 Philippines
}%




\date{\today}

\begin{abstract}
The weak equivalence principle (WEP) in the quantum regime has been the subject of many studies with a broad range of approaches to the problem. Here we tackle the problem anew through the time of arrival (TOA) operator approach.  This is done by constructing the TOA operator for a non-relativistic and structureless particle that is projected upward in a uniform gravitational field with an intended arrival point below the classical turning point. The TOA operator is constructed under the constraint that the inertial and gravitational masses are equivalent, and that Galilean invariance is preserved. These constraints are implemented by Weyl quantization of the corresponding classical TOA function for the projectile. The expectation value of the TOA operator is explicitly shown to be equal to the classical time of arrival plus mass dependent quantum correction terms, implying incompatibility of the WEP with quantum mechanics. The full extent of the violation of the WEP is shown through the mass dependence of TOA distribution for the projectile. 
\end{abstract}

\pacs{03.65.Ta, 04.20.Cv}
\maketitle


\section{Introduction}
\label{sec:intro}

Space and time are the basic entities in physics which provide the framework for any description of natural processes \cite{toa2a,toa2b}. Despite this, quantum mechanics and general relativity, which are the most successful theories we have to date, have fundamentally different concepts of time. Quantum mechanics merely treats time as an external parameter which governs the evolution of the state of the system. Meanwhile, time in general relativity is dynamical, and its dynamics are influenced by the geometry of spacetime which allows material clocks to display proper time. Furthermore, these clocks react to the metric changing the geometry \cite{macias,zeh}. These different treatments of time in quantum quantum mechanics and general relativity then pose a problem as quantum mechanics breaks down when considering quantum phenomena that interact with the background spacetime \cite{macias}. To reconcile the two, a possible option may be to either introduce a non-dynamical time in general relativity or a dynamical time in quantum mechanics \cite{macias}. Here, we take the latter route, i.e. we introduce a time of arrival (TOA) operator to accommodate the concept of a dynamical time in quantum mechanics. With this, the compatibility of general relativity and quantum mechanics can be studied through the equivalence principle. Specifically, we consider the simplest case where a non-relativistic and structureless particle, with an initial velocity $v=v_o$ and initial position $q=-q_o$, is projected upward in a uniform gravitational field and study if the equivalence principle in its weak form remains valid for the quantum image of such a system.

The equivalence principle played a central role in the development of general relativity, and has various formulations associated with different physical meanings \cite{epstatement}. In classical mechanics, the equivalence principle can be expressed by three equivalent but physically distinct statements: ({\it i}) the equivalence of inertial and gravitational masses, ({\it ii}) the equivalence of a state of rest in a homogeneous gravitational field and the state of uniform acceleration in the absence of gravity, and ({\it iii}) the equivalence of motion for all sufficiently small bodies in free fall, i.e. all bodies fall with the same acceleration independent of their composition and mass. The third statement is known as the weak equivalence principle (WEP). Since each statement implies the other two, it only takes one of them to be confirmed to establish the validity of the other statements, as well as the geometrical nature of gravity. Any experiment that tests the equivalence principle for classical systems (EPCS) serves as a test of the foundations of general relativity that can lead to a search for a new long range field coupling to matter that depends on composition \cite{experimenta}. The equivalence principle is so fundamental that whether a violation can be confirmed or not, both results hold great significance. A confirmed violation of the equivalence principle is as significant as discovering a new fundamental force of nature. Meanwhile, the latter case can push the limits of current experimental techniques to improve the accuracy of the equivalence principle. Several modern experiments have been performed to test the equivalence principle using different techniques such as: ({\it i}) the use of a torsion balance (mainly due to E\"{o}tv\"{o}s \cite{eotvosa,eotvosb}), ({\it ii}) using the Sun as a daily modulated signal source \cite{sun1,sun2,sun3}, ({\it iii}) rotating torsion balance \cite{wagner} (and references therein), and ({\it iv}) lunar laser ranging \cite{lunarlaser1,lunarlaser2,lunarlaser3}. To date, preliminary results of the Microscope satellite have validated the WEP at the accuracy of $10^{-14}$ for the titanium-platinum E\"{o}tv\"{o}s parameter \cite{wepexp}. It is expected that the accuracy should reach $10^{-15}$ when all the data over the whole lifetime of the satellite are analyzed. (Ref. \cite{experimenta} provides a more comprehensive discussion of the current progress done to further improve the accuracy of the equivalence principle).

The geometrical nature of gravity in classical physics has been well established. Now, since quantum laws are supposedly the fundamental laws of nature, we inquire into the quantum status of the geometrical nature of gravity. Quantum mechanics treats gravity on equal footing with the rest of the forces. Moreover, the quantum equation of motion, specifically the Schr\"{o}dinger equation, explicitly depends on the mass of the object. These seem to indicate that the equivalence principle is not compatible with quantum systems. However, the experiment done by Colella, Overhauser and Werner has confirmed the validity of the first statement of the EPCS, while the experiment done by Bonse and Wroblewski has confirmed the validity of the second statement \cite{cow_wrob1,cow_wrob2}. This now leaves the third statement, i.e the WEP, open to further study for quantum systems. 

Several theoretical investigations of the WEP in the quantum regime have been made, which cover a broad range of approaches to the problem \cite{daviesfang,candelassciama,alvarezmann,dalvit,singleton,greenberger,zych_brukner,orlando,mousavi2015,chowdhury2012,viola1997,davies2004a,ali2006,statement}. Some of the approaches to the problem that treat time merely as a parameter in quantum mechanics are Refs. \cite{viola1997,davies2004a,ali2006}. Viola and Onofrio proposed to test the equivalence principle through freely falling quantum objects using Ehrenfest theorem to compute for the average time of flight (TOF) of a falling quantum particle. They found that due to the linearity of the potential, the average TOF equals that of the classical TOF \cite{viola1997}. However, there was mass dependence on the width of the TOF distribution. The same problem was treated by Davies using a quantum clock analysis, specifically a Peres clock, to the motion of a quantum particle in a stationary state in a gravitational field \cite{davies2004a}. He showed that there is a mass dependent quantum correction term to the classical transit time due to tunneling. Nonetheless, the quantum transit time approaches the classical transit time for points far from the classical turning point. He took this as a quantum manifestation of the weak equivalence principle. Ali et. al. treated the same problem via the Bohmian-trajectory approach and showed that there is mass dependence on both the position detection probabilities, and the mean arrival time \cite{ali2006}. The WEP was then recovered in the limit of large mass. 

It can thus be seen from the previous studies mentioned above that a violation of the WEP is always present for quantum systems, and can thus be concluded that the WEP does not have a quantum analogue, i.e. the WEP and quantum theory are incompatible with each other. However, it has been argued in Ref. \cite{statement} that the violation of the WEP for quantum systems may as well be a consequence of the fundamentally opposing realities of classical mechanics and quantum mechanics, i.e. classical mechanics is deterministic while quantum mechanics is probabilistic. A statement of the WEP for quantum systems should be formulated solely from quantum concepts with no reference to classical constructs. Two of the main arguments are: ({\it i}) the concept of a trajectory in quantum mechanics is not well-defined due to the non-locality of the particle, and ({\it i}) the mass does not cancel from the time evolution of quantum states \cite{zych_brukner,statement}. With these considerations, a statement of the WEP for quantum systems has been proposed in Ref. \cite{statement} which we shall call the Anastopoulos-Hu WEP for quantum systems (AHWEP). The AHWEP can be expressed by two operationally distinct statements which should apply to all quantum states even to those without a classical analogue: ({\it i}) the probability distribution of position for a free-falling particle is the same as the probability distribution of a free particle, modulo a mass-independent shift of its mean, and ({\it ii}) any two particles with the same velocity wavefunction behave identically in free fall. The second statement of the AHWEP then implies that the TOA distribution of two quantum particles with the same initial group velocity $v_o$ should be identical regardless of their mass. It follows that if we consider a structureless quantum particle with initial group velocity $v=v_o$ and initial mean position $q=-q_o$ that is projected upward in a uniform gravitational field, then the TOA distribution at the arrival point should be identical regardless of mass. 

Standard quantum mechanics postulates that the probability distribution of the measurement outcomes of an observable can be obtained from the corresponding operator representation of the observable. This then means that to construct the TOA distribution, we need to introduce a TOA operator in quantum mechanics. However, the incorporation of time as a dynamical observable in quantum mechanics is widely known as the quantum time problem, which involves the introduction of a Hermitian time operator that is canonically conjugate to the system Hamiltonian. The existence of this time operator has been opposed by Pauli, which led many researchers to introduce a time operator with a compromise, i.e. either give up Hermiticity or conjugacy with the system Hamiltonian \cite{toa2a,toa2b}. Nonetheless, one of us has demonstrated that Pauli's objection does not hold in the Hilbert space formulation of quantum mechanics, and has proved the existence of a Hermitian time operator that is canonically conjugate with the system Hamiltonian \cite{G2002PRSLA1_a,G2002PRSLA1_b,GCB2004PRL_a,GCB2004PRL_b}. Here, we tackle the WEP for quantum systems anew using the theory of quantum TOA operators advocated in Refs. \cite{G2002PRSLA1_a,G2002PRSLA1_b}, which leads to the introduction of a dynamical time in quantum mechanics. We emphasize that our calculations will only involve non-relativistic quantum mechanics in the weak gravity regime to avoid possible complications that may arise when considering relativistic particles. Depending on the energy of the particle, there is a possibility that particle creation and annihilation will occur. The concept of time of arrival then loses meaning since we are not sure if the particle that arrived is the same particle that was fired from the initial point. 

The rest of the paper is organized as follows. The TOA operator is constructed by quantizing the classical time of arrival using Weyl quantization in Sec. \ref{sec:quantization}. The expectation value of the TOA operator, $\tau_{\text{quant}}$, for an arbitrary single-peaked wavepacket is then calculated in Sec. \ref{sec:toa_expec}. The classical TOA is then extracted from $\tau_{\text{quant}}$, and shown that there are mass-dependent quantum correction terms to the classical TOA. A Gaussian wavepacket is used as an example to explicitly calculate the quantum correction terms in Sec. \ref{sec:quant_corr}. The TOA distribution for these Gaussian wavepackets are then constructed in Sec. \ref{sec:toa_dist}, and shown to be mass-dependent. Lastly, Sec. \ref{sec:conc} summarizes the paper and concludes.

\section{Quantizing the classical time of arrival}
\label{sec:quantization}

First, let us consider the free fall motion classically. We start by imposing the first statement of the EPCS, i.e. the equivalence of the inertial and gravitational masses $m_i=m_g=\mu$. The TOA at the origin for a classical point particle with initial velocity $v=v_0$ and position $q=-q_0$ that is projected upward is given by
\begin{equation}
T_{\pm}=\frac{v_o}{g} \left( 1 \pm \sqrt{1 -  \frac{2g q_o}{v_o^2}} \right).
\label{falltime0}
\end{equation}
The negative sign corresponds to the case when the motion of the particle is restricted to one time crossing at the origin. On the other hand, the positive sign corresponds to the case where the particle can cross the origin twice, e.g. the particle is projected upward and reaches its maximum height then moves down until it crosses the origin.  The time of arrival $T$ is complex when $(2 g  q_o/v_o^2)>1$, indicating non-arrival at the origin. The classical TOA is independent of mass, which is a statement of the WEP, i.e. bodies fall with the same acceleration regardless of their composition and their masses.

Quantum mechanically, we do not expect that an ensemble of particles prepared in the same initial state will arrive at the origin at the same time. We then get a distribution of the TOA at the arrival point. If we expect that the WEP is also true for quantum systems, then following from AHWEP, the TOA distribution of two particles with the same initial group velocity and mean position should be identical, regardless of composition and mass. If there is a difference in the TOA distribution, then the particle can be identified using its TOA distribution. This implies that the WEP and quantum theory are still incompatible despite both the introduction of a dynamical time in quantum mechanics, and formulating a statement of the WEP solely from quantum concepts.

In standard quantum mechanics, the distribution of the measurement outcomes of a quantum observable is constructed using the spectral resolution of the operator corresponding to the observable. Naturally, to address the quantum time of arrival problem within standard quantum mechanics, one needs to construct the appropriate TOA operator $\opr{T}$ that is canonically conjugate with the system Hamiltonian. However, the consensus is that no such operator can be constructed in the most general case of arbitrary arrival point and of arbitrary interaction potential. In one dimension, the most quoted reason is that the classical TOA at any given point does not admit a sensible quantization because it is generally not everywhere real and single valued in the entire phase space \cite{jaykel,reason1,reason2}. These problems are evident in Eq. \eqref{falltime0}. Nonetheless, the problem of quantizing the classical time of arrival observable has already been addressed in Ref. \cite{jaykel}.

We now deal with the construction of the quantum TOA operator corresponding to the classical time of arrival Eq. \eqref{falltime0}. The classical TOA Eq. \eqref{falltime0} is multiple valued but it is only reasonable to quantize the first TOA. This is a physical constraint arising from the fundamental difference in the nature of classical and quantum mechanics. For classical systems, we can perform a measurement without disturbing the system. However, in quantum mechanics, performing a measurement induces an irreversible change to the state of the system. This means that the state of the system after recording the first TOA is no longer causally related to the state of the system before the measurement. Moreover, the second TOA can no longer be interpreted as the second TOA starting from the initial state \cite{jaykel}. In the following calculations we will only quantize $T_{-}$ in Eq. \eqref{falltime0} which corresponds to the first time crossing at the origin.

We proceed through quantization by first rewriting Eq. \eqref{falltime0} into a form amenable to quantization. The initial initial velocity $v_0$ is expressed in terms of the initial momentum $p_0=\mu v_0$. Imposing that the classical TOA should be real and single-valued, we expand Eq. \eqref{falltime0} in binomial series. By doing so, we arrive at the following expansion of $T_-$, 
\begin{equation}
T= 2 \mu \sum_{n=0}^{\infty} \binom{1/2}{n+1} (-2 \mu^2 g)^n \frac{q_0^{n+1}}{p_0^{2n+1}} \label{eqltoa},
\end{equation}
which only converges when the initial kinetic energy is greater than the potential energy at the arrival point. That is, the particle still continues to move upward after it reaches the arrival point. 

In standard quantum mechanics, the second statement of the EPCS, i.e. equivalence of a state of rest in a homogeneous gravitational field and the state of uniform acceleration in the absence of gravity, is already embedded in the Schrodinger equation and can be done by performing a coordinate transform. With the introduction of a TOA operator in quantum mechanics, the second statement of the EPCS can be used to determine the quantization rule to be used in quantizing the classical TOA. Recall that for any operator $\opr{A}$ the equation of motion of the operator $\opr{A}$ in the Heisenberg representation is given by
\begin{equation}
\frac{d \opr{A}}{dt} = \frac{i}{\hbar}[\opr{H},\opr{A}]+\frac{\partial \opr{A}}{\partial t},
\label{eqmotionquant}
\end{equation}
where $[\opr{H},\opr{A}]$ is the commutator of the Hamiltonian $\opr{H}$ with the operator $\opr{A}$. The classical analogue of Eq. \eqref{eqmotionquant} is
\begin{equation}
\frac{d a}{dt} = \{a,H\}+\frac{\partial a}{\partial t},
\label{eqmotionclass}
\end{equation}
where $\{a,H\}$ is the Poisson bracket of the classical observable $a$ corresponding to the operator $\opr{A}$ with the Hamiltonian. Imposing the second statement of the EPCS then means that Eq. \eqref{eqmotionclass} must be equivalent with its quantum analogue Eq. \eqref{eqmotionquant}. Now, if the operator $\opr{A}$ is the TOA operator $\opr{T}$ this leads us to quantize Eq. \eqref{eqltoa} under the condition that the time-Hamiltonian Poisson bracket goes over to the canonical commutator relation: $\{H,T\}=1\rightarrow [\opr{H},\opr{T}]=i\hbar\opr{I}$. This restricts quantization by Weyl quantization of $T$, which yields
\begin{equation}
\opr{T}= \mu \sum_{n=0}^{\infty} \binom{1/2}{n+1} (-\mu^2 g)^{n} \sum_{k=0}^{n+1}\opr{q}^k \opr{p}^{-2n-1} \opr{q}^{n+1-k}.
\label{quantoa}
\end{equation}

In coordinate representation, the time of arrival operator is the integral operator $(\opr{T}\phi)(q)=\int_{-\infty}^{\infty} \kernel{\opr{T}}\phi(q') dq'$, where the kernel is given by
\begin{align}
\kernel{\opr{T}}=& 2 \mu \sum_{n=0}^{\infty} \binom{1/2}{n+1} (-2 \mu^2 g)^n  \frac{1}{2^{n+1}} \nonumber \\
& \times \sum_{k=0}^{n+1} \binom{n+1}{k} q^k q'^{n+1-k} \kernel{\opr{p}^{-2n-1}}. 
\end{align}
Performing the summation, the kernel assumes the form 
\begin{align}
\kernel{\opr{T}}= &\left(\frac{\mu^2 g}{\hbar^2}(q+q')(q-q')^2\right)^{-1/2} \nonumber \\
& \times J_1\left(\sqrt{\frac{\mu^2 g}{\hbar^2}(q+q')(q-q')^2}\right) \nonumber \\
&\times \frac{\mu i}{\hbar} \frac{(q+q')}{2}\mbox{sgn}(q-q'),
\label{kernel}  
\end{align}
where we used the identity \cite{jaykel,table}
\begin{equation*}
\kernel{\opr{p}^{-m}} = \frac{i}{2} \dfrac{(-1)^{(m-1)/2}}{\hbar^m (m-1)!} (q-q')^{m-1} \mbox{sgn}(q-q').
\end{equation*}
Moreover, $\mbox{sgn}(z)$ is the sign function and $J_1(z)$ is a particular Bessel function of the first kind. The singularity of the kernel along the diagonal $q=q'$ can be removed using the identity $\sqrt{z}^{-1}J_{1}(\sqrt{z})=2^{-1}{_0 F_1}(;2;-z/4)$ where ${_0 F_1}(;a;z)$ is a particular hypergeometric function.

We claim that the quantized TOA-operator $\opr{T}$ is a legitimate quantum first time of arrival operator by virtue of the dynamics of its eigenfunctions. The eigenfunctions unitarily evolve through time to localize at the intended arrival point  at their corresponding eigenvalues, a phenomenon we referred to as unitary arrival \cite{GCB2004PRL_a,GCB2004PRL_b,denny,denny2,jaykel}. 
The eigenfunctions fall under two kinds--- non-nodal and nodal eigenfunctions. The former has the characteristic dynamical property that a single peak gathers at the arrival point with its minimum width occurring at its eigenvalue, and it corresponds to particle arrival with detection. On the other hand, the later has the characteristic dynamical property that two peaks gather at the arrival point with their closest separation also occurring at its eigenvalue, and it corresponds to particle arrival without detection \cite{jaykel,denny,denny2}. A pair of nodal and non-nodal evolving eigenfunctions are shown in Fig. \ref{fig:toaeigenfuncevo}. (See Appendix-B for a discussion in solving the eigenvalue problem for the time of arrival operator $\opr{T}$.)

\begin{figure*}[t]
\begin{subfigure}
\centering
\includegraphics[width=0.475\textwidth,height=5cm]{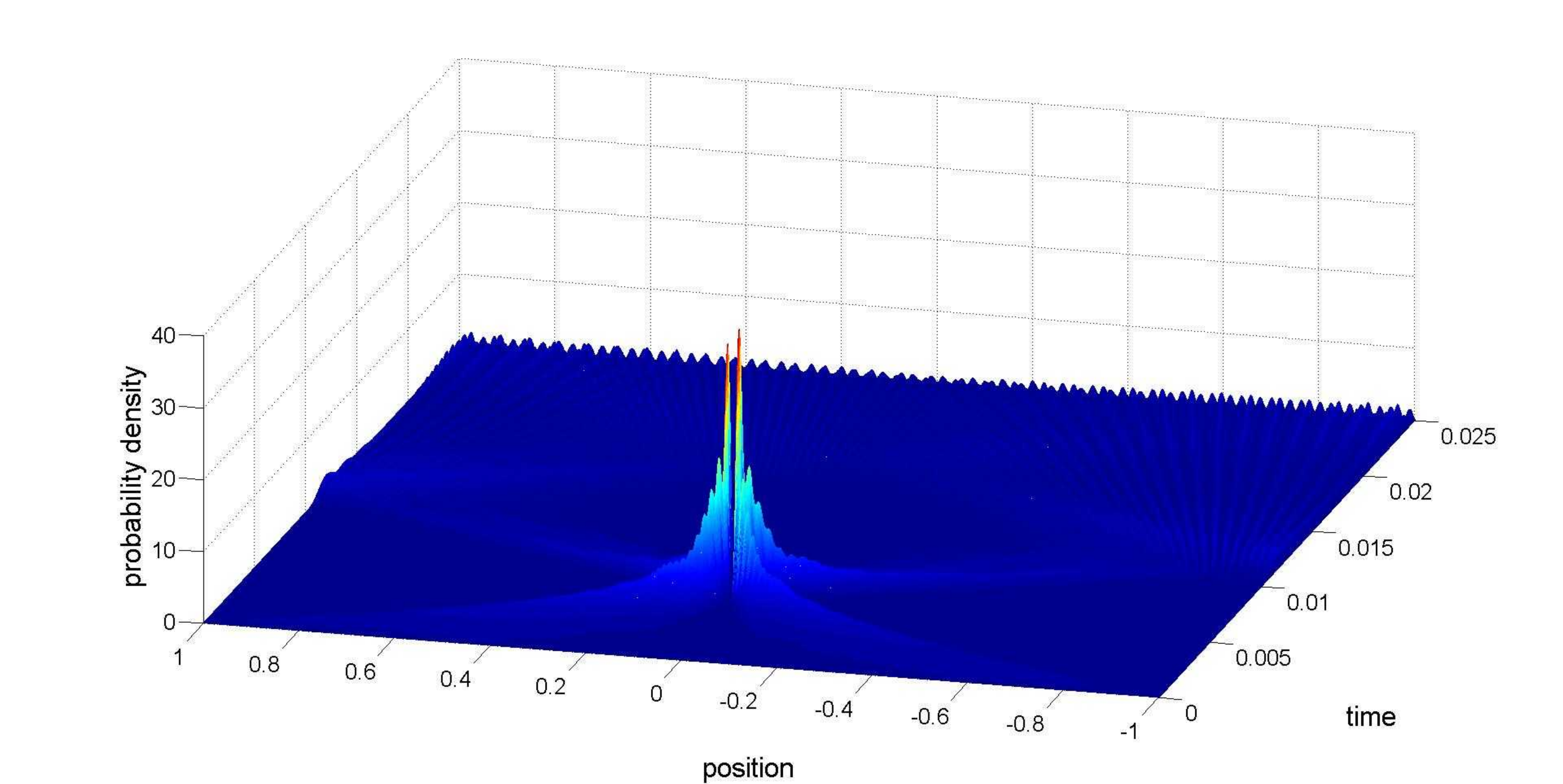}
\end{subfigure}
\begin{subfigure}
\centering
\includegraphics[width=0.475\textwidth,height=5cm]{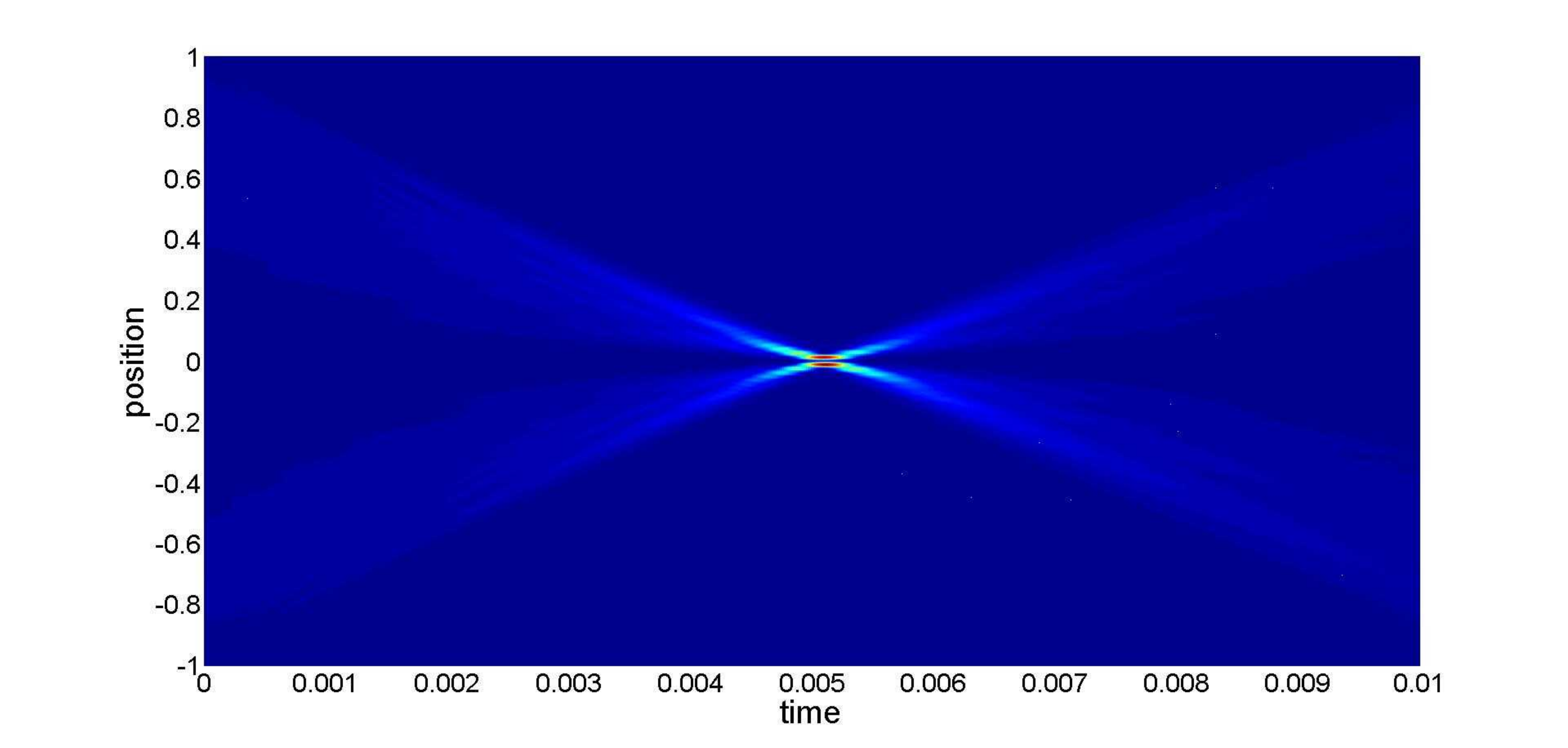}
\end{subfigure}
\begin{subfigure}
\centering
\includegraphics[width=0.475\textwidth,height=5cm]{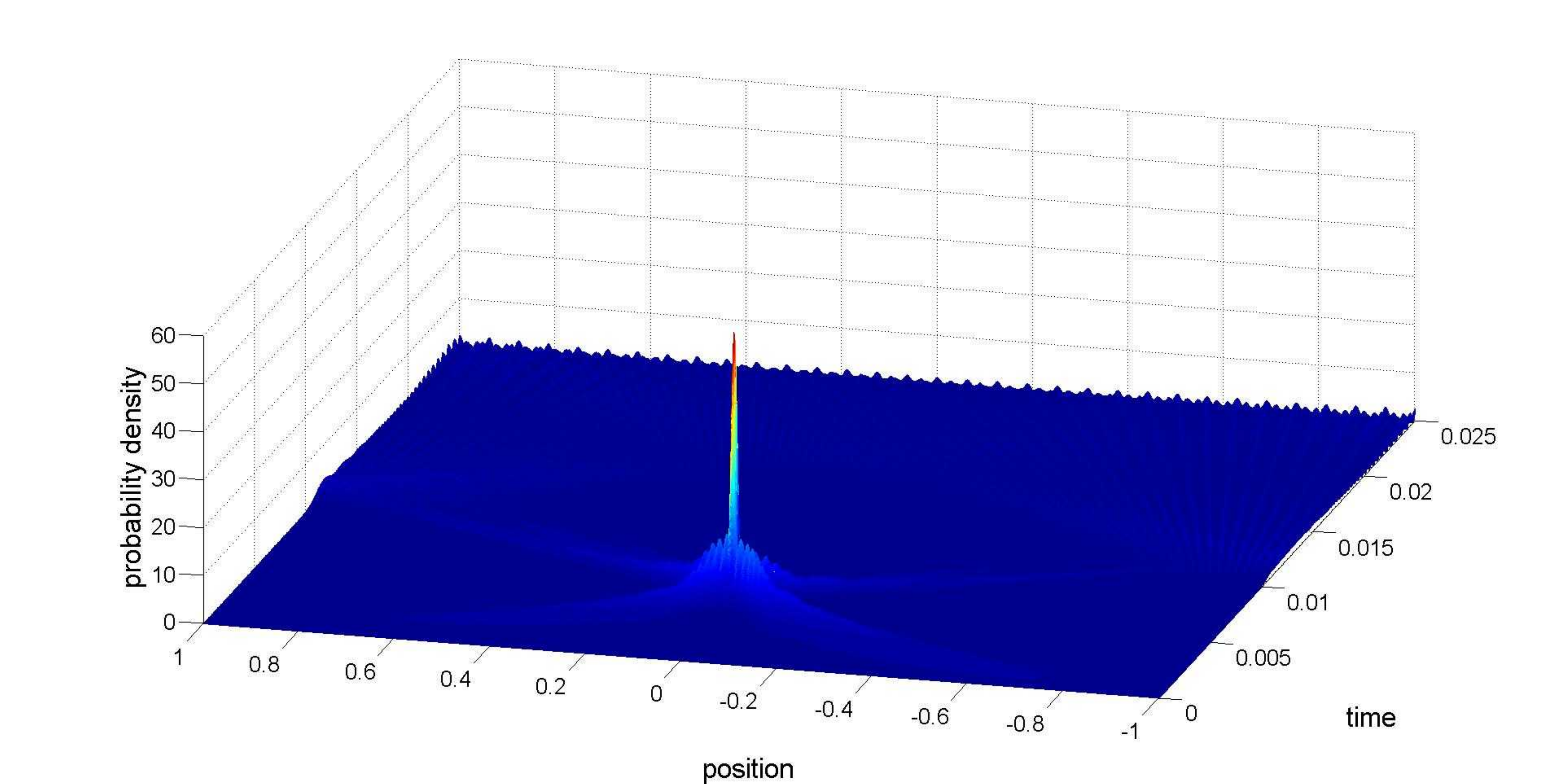}
\end{subfigure}
\begin{subfigure}
\centering
\includegraphics[width=0.475\textwidth,height=5cm]{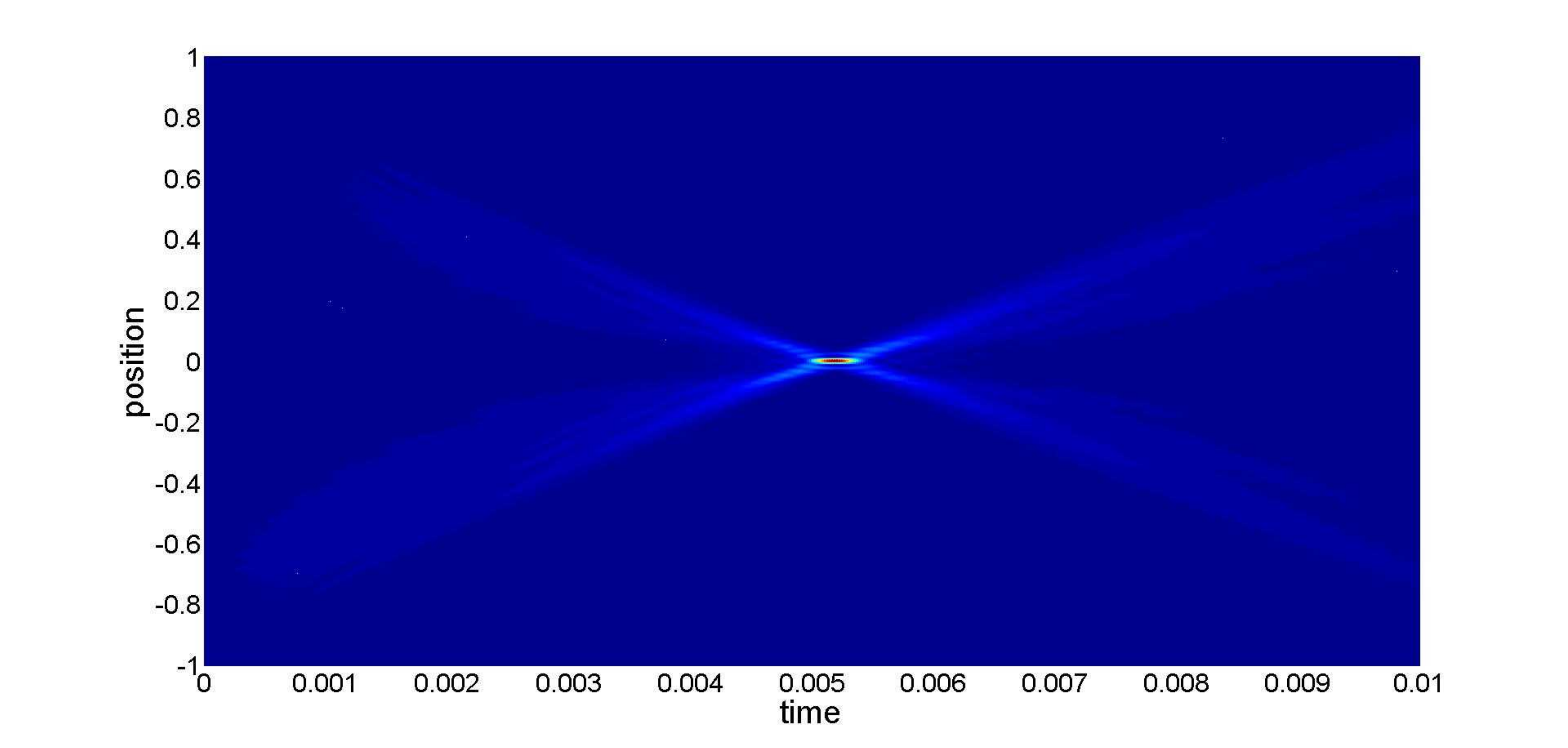}
\end{subfigure}
\caption{Time evolution of the nodal (top) and non-nodal (bottom) eigenfunctions of the TOA operator with eigenvalues $0.00508$ and $0.00517$, respectively, for the parameters $\mu=g=\hbar=1$.}
\label{fig:toaeigenfuncevo}
\end{figure*}

\section{Time of arrival for single particles}
\label{sec:toa_expec}

We now consider the expected time of arrival at the origin for a quantum projectile with mass $\mu$ and with initial upward group velocity $v_0$. We take the expected time of arrival to be equal to the average of an arbitrary large number of independent measurements of the time of arrival of the projectile at the origin. We assume that the projectile is prepared in a pure state $\phi(q)$, which leads to an initial wavefunction of the form $\phi(q)=e^{i\mu v_o q/\hbar}\varphi(q)$. The wavepacket $\varphi(q)$ satisfies the property $\langle\varphi|\opr{p}|\varphi\rangle=0$, where we assume that $\varphi(q)$ is independent of $\hbar$. The expected time of arrival is now postulated to be equal to the expectation value of the time of arrival operator $\opr{T}$,  $\bar{\tau}_{quant}=\langle\phi|\opr{T}|\phi\rangle=\int_{-\infty}^{\infty}\bar{\phi}(q)(\opr{T}\phi)(q)dq$, which is explicitly given by
\begin{align}
	\bar{\tau}_{\text{quant}}=& \frac{\mu i}{\hbar} \int_{-\infty}^{\infty} \int_{-\infty}^{\infty} \phi^*(q)   \left(\frac{\mu^2 g}{\hbar^2} (q-q')^2 (q+q')\right)^{-1/2} \nonumber \\
	&\times \left(\frac{q+q'}{2}\right) J_1\left(\sqrt{\frac{\mu^2 g}{\hbar^2} (q-q')^2 (q+q')}\right) \nonumber \\
	& \times  \text{sgn}(q-q') \phi(q') dq'dq.
	\label{toaexpec}
\end{align}
While the classical time of arrival at the origin can be complex, the quantum expected time of arrival is real valued for all initial wavefunctions $\phi(q)$, taking its values in the entire real line. For sufficiently localized wavepackets projected upward, the expected time of arrival is finite and positive. 

Eq. \eqref{toaexpec} already shows mass dependence of the expected time of arrival. This signals departure from the WEP because the expected arrival time can be used to distinguish projectiles with different masses. However, it may happen that the mass dependence of the incident wavepacket, $\phi(q)=e^{i\mu v_o q/\hbar}\varphi(q)$, cancels the mass dependence of the time of arrival operator, in much the same way that mass dependence cancels out in the classical case. From the quantum-classical correspondence principle, we expect that the classical time of arrival must emerge from equation \eqref{toaexpec} in the classical limit, $\hbar\rightarrow 0$. Since the classical time of arrival is already independent of mass, any departure of the quantum expected time of arrival from the classical time of arrival must necessarily involve corrections that depend on mass from mere dimensional analysis. We now confirm this by obtaining an expansion of the expectation value of $\opr{T}$ in powers of $\hbar$. To accomplish this, we perform the change of variables, $q=x+y$ and $q'=x-y$, to cast Eq. \eqref{toaexpec} in the form
\begin{align}
\bar{\tau}_{\text{quant}} =& \frac{\mu i}{\hbar} \int_{-\infty}^{\infty}\int_{-\infty}^{\infty} xe^{-2i\mu v_o y/ \hbar} {_0 F_1}\left(;2;-\frac{2 \mu^2 g}{\hbar^2}xy^2\right) \nonumber \\
&\times \bar{\varphi}(x+y)\varphi(x-y)\text{sgn}(y)dx dy,
\label{toaexpec1}
\end{align}
where we have written the Bessel function in terms of a hypergeometric function. 

To proceed, we perform a Taylor series expansion on $\bar{\varphi}(x + y)\varphi(x-y)$ and ${_0 F_1}(;2;-2 \mu^2 gxy^2/\hbar^2)$ about $y=0$. The order of summation and integration is then interchanged to separate the integrals over $x$ and $y$. The resulting integrals in $y$ are divergent and are interpreted as a  distributional Fourier transform \cite{table}. The relevant integrals are given by
\begin{equation*}
\int_{-\infty}^{\infty}y^{m-1}e^{-i \nu y}\text{sgn}(y)dy=\dfrac{2(m-1)!}{(i\nu)^m},
\end{equation*}
for $m=1, 2,\dots$. Performing the indicated operations and rearranging the order of the summation to collect like powers of $\hbar$, Eq. \eqref{toaexpec1} assumes the expansion
\begin{align}
\bar{\tau}_{\text{quant}}=& \frac{1}{\sqrt{\pi}v_o} \sum_{r=0}^{\infty}\left(\frac{i \hbar}{\mu v_o}\right)^r \dfrac{\Gamma(\frac{r+1}{2})\Gamma(\frac{r+2}{2})}{r!} \nonumber \\
& \times \int_{-\infty}^{\infty} {_2 F_1}\left(\frac{r+1}{2},\frac{r+2}{2};2;\frac{2g}{v_o^2}x\right) x W_r(x) dx
\label{toaexpecfinal}
\end{align}
where, 
\begin{equation}
W_r(x) = \sum_{q=0}^r \dfrac{r!}{(r-q)!q!} (-1)^q \bar{\varphi}^{(q)}(x)\varphi^{(r-q)}(x).
\label{quantcorr}
\end{equation}

Eq. \eqref{toaexpecfinal} has two important properties. First, the series is generally divergent. However, meaningful numerical results can be obtained by interpreting the series as an asymptotic expansion of Eq. \eqref{toaexpec} for small values of the parameter $(\hbar/\mu v_0)$. This implies that Eq. \eqref{toaexpecfinal} is only valid for either large values of mass or large values of initial speed $v_0$. Thus, it describes the behavior of the expectation value in the semiclassical regime. Second, while the expansion follows from a real valued expression, each term of the expansion may be complex when the support of $\varphi(q)$ is sufficiently large. This follows from the fact that the hypergeometric function $ _2F_1(a,b;c;z)$ has a branch cut along $[1,\infty)$. The terms become complex when the integration extends beyond the branch point. The emergence of complex values for the expected time of arrival is related to the phenomenon of missing terms when interchanging the order of integration and summation lead to divergent integrals \cite{wongBook,wongPAMS1980,galaponPRSA2017, tica2019, tica2018}. The divergent integrals signal the presence of terms that are missed out by, in our present case, interpreting the divergent integrals as distributional Fourier transforms. The missing terms should be responsible for the cancellation of the imaginary part of the complex terms in the expansion to maintain the real valuedness of the expected time of arrival. However, it is beyond the scope of the paper to treat the problem of missing terms in our expansion \eqref{toaexpecfinal}. It will be sufficient for us to physically motivate our expansion and confirm its numerical accuracy against the exact value given by Eq. \eqref{toaexpec} in our subsequent discussions. 

To make sense of the complexity of the expansion, we look into how the classical time of arrival emerges from the expansion \eqref{toaexpecfinal}. It must emerge from the term independent of $\hbar$, which is the leading term, 
\begin{align}
\tau_{0}=&\dfrac{1}{v_o}\int_{-\infty}^{\infty}{_2 F_1}\left(\frac{1}{2},1;2;\frac{2 g}{v_o^2}x\right)x|\varphi(x)|^2 dx.
\label{term0}
\end{align}
Using the identity ${_2 F_1}(a,a+1/2;2a+1;z)=2^{2a}(1+\sqrt{1-z})^{-2a}$, we obtain
\begin{equation}
    \frac{x}{v_0} \, _2F_1\left(\frac{1}{2},1;2;\frac{2g}{v_0^2}x\right) =\dfrac{v_o}{g} \left(1 - \sqrt{1-\dfrac{2 g}{v_o^2}x} \right),\label{ctoa}
\end{equation}
which we recognize as the classical time of arrival at the origin. Then, $\tau_0$ is just the expectation value of the classical time of arrival, where the initial launching point $x$ is drawn from the distribution $|\varphi(x)|^2$. Clearly $\tau_o$ is complex when $2gx/v_o^2>1$, that is when the corresponding classical particle has no sufficient energy to arrive at the origin. Since the hypergeometric functions in Eq. \eqref{toaexpecfinal} have common branch points at $2gx/v_0^2=1$, all terms in the expansion \eqref{toaexpecfinal} are complex when the leading term $\tau_0$ is complex. Since the complexity of $\tau_0$ arises from the corresponding classical particle not having sufficient energy to reach the arrival point, the group of missing terms in the expansion encapsulates the quantum tunneling effect which is not captured by the classical time of arrival expression  \eqref{ctoa}. Then the expansion in equation \eqref{toaexpecfinal} is a meaningful semiclassical expansion provided quantum tunneling is negligible. This condition is satisfied if the incident wavepacket has a spread or support that is sufficiently small such that $2g x/v_o^2<1$ for all $x$ in the support of $\varphi(q)$. In the rest of the paper, we assume that such condition is satisfied or at most the initial wavefunction is exponentially suppressed in the classically forbidden region. 
\begin{figure}[t]
	\centering
	\includegraphics[width = 0.4\textwidth]{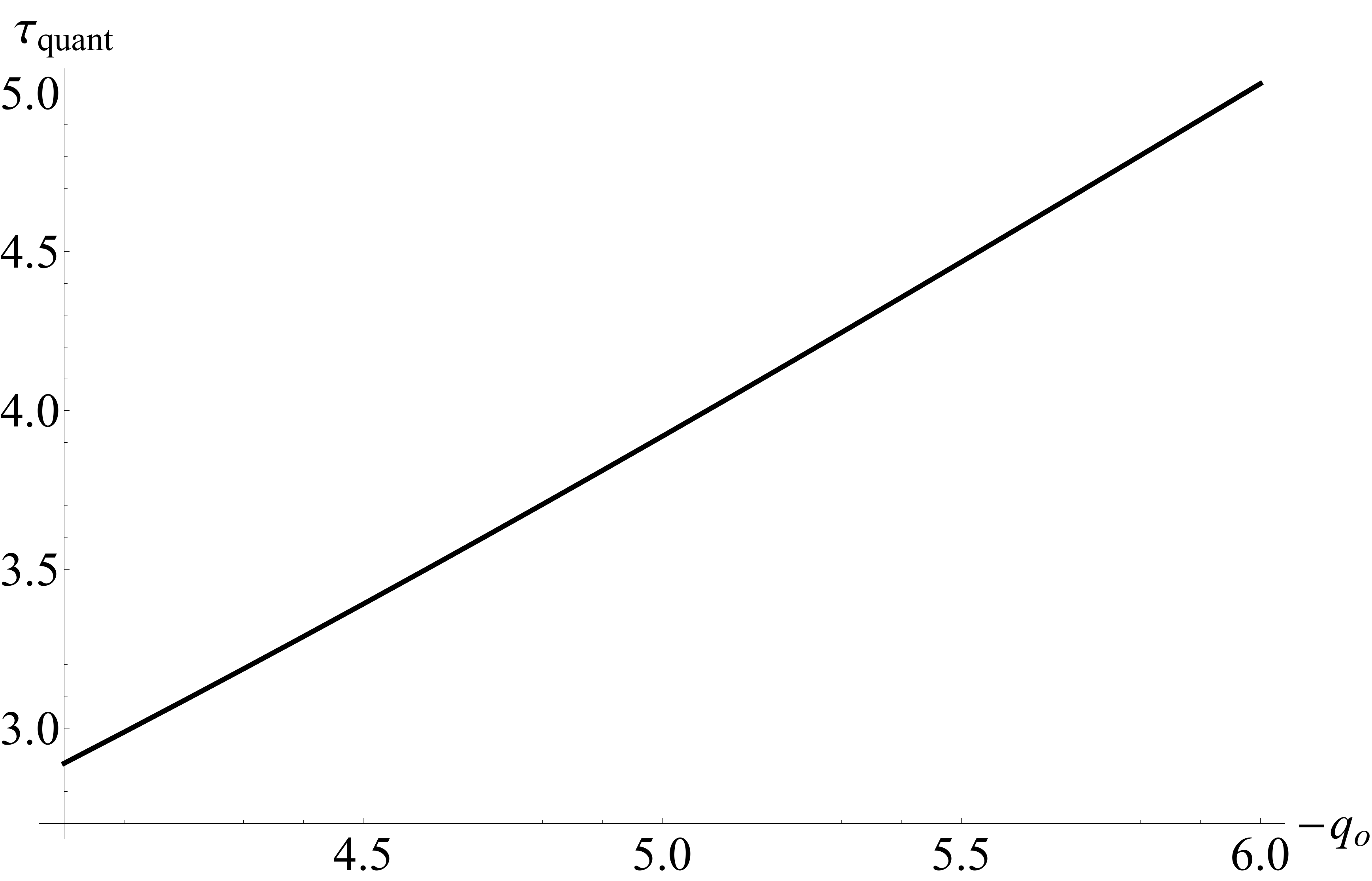}
	\caption{Quantum time of arrival at the origin for a single atom described as a Gaussian wavepacket where the classical turning point is below the origin for the parameters $\mu=g=\hbar=1$, initial velocity $v_o=2$, and $\sigma^2=0.1$ for varying intial position $q_o$.}
	\label{fig:tunnel}
\end{figure}

Now, if $\varphi(q)$ is single peaked and localized around $x=q_o$, then we have
\begin{equation}
\tau_{0}=\dfrac{v_o}{g} \left(1 - \sqrt{1-\dfrac{2 g}{v_o^2}q_o} \right),
\end{equation}
which is equal to the classical value defined in Eq. \eqref{falltime0}. The subsequent terms in the expansion \eqref{toaexpecfinal} are in positive powers of $\hbar$ so that they represent quantum corrections to the classical time of arrival. We can then rewrite the expansion in the more transparent form,
\begin{equation*}
    \bar{\tau}_{\text{quant}} = \tau_{class} + \sum_{r=1} \alpha_r \hbar^r
\end{equation*}
where 
\begin{align}
\alpha_r =& \frac{1}{\sqrt{\pi}v_o} \left(\frac{i }{\mu v_o}\right)^r \dfrac{\Gamma(\frac{r+1}{2})\Gamma(\frac{r+2}{2})}{r!} \nonumber \\
&\times \int_{-\infty}^{\infty} x {_2 F_1}\left(\frac{r+1}{2},\frac{r+2}{2};2;\frac{2g}{v_o^2}x\right) W_r(x) dx.
\end{align}
The presence of these quantum correction terms already imply that WEP and quantum theory are incompatible with each other. The magnitude of these quantum correction terms depends on the initial state of the wavefunction (e.g. mass, velocity, spread of the wavepacket) and it can be seen that despite imposing the validity of the first and second statement of the EPCS to be carried over to quantum systems, a violation of the WEP for quantum systems still arises. These quantum correction terms may be positive, negative, or zero depending on the initial state, with the first two corresponding to an advanced and a delayed arrival of the particle, respectively. Then using only the classical TOA to describe the TOA of the particle is insufficient to describe the total TOF of the particle. However, the effect of these quantum correction terms can be minimized by introducing an appropriate position-dependent phase on the initial wavefunction that can make the quantum correction terms vanish up to a certain order \cite{self}.

\begin{figure}[t]
	\centering
	\includegraphics[width = 0.4\textwidth]{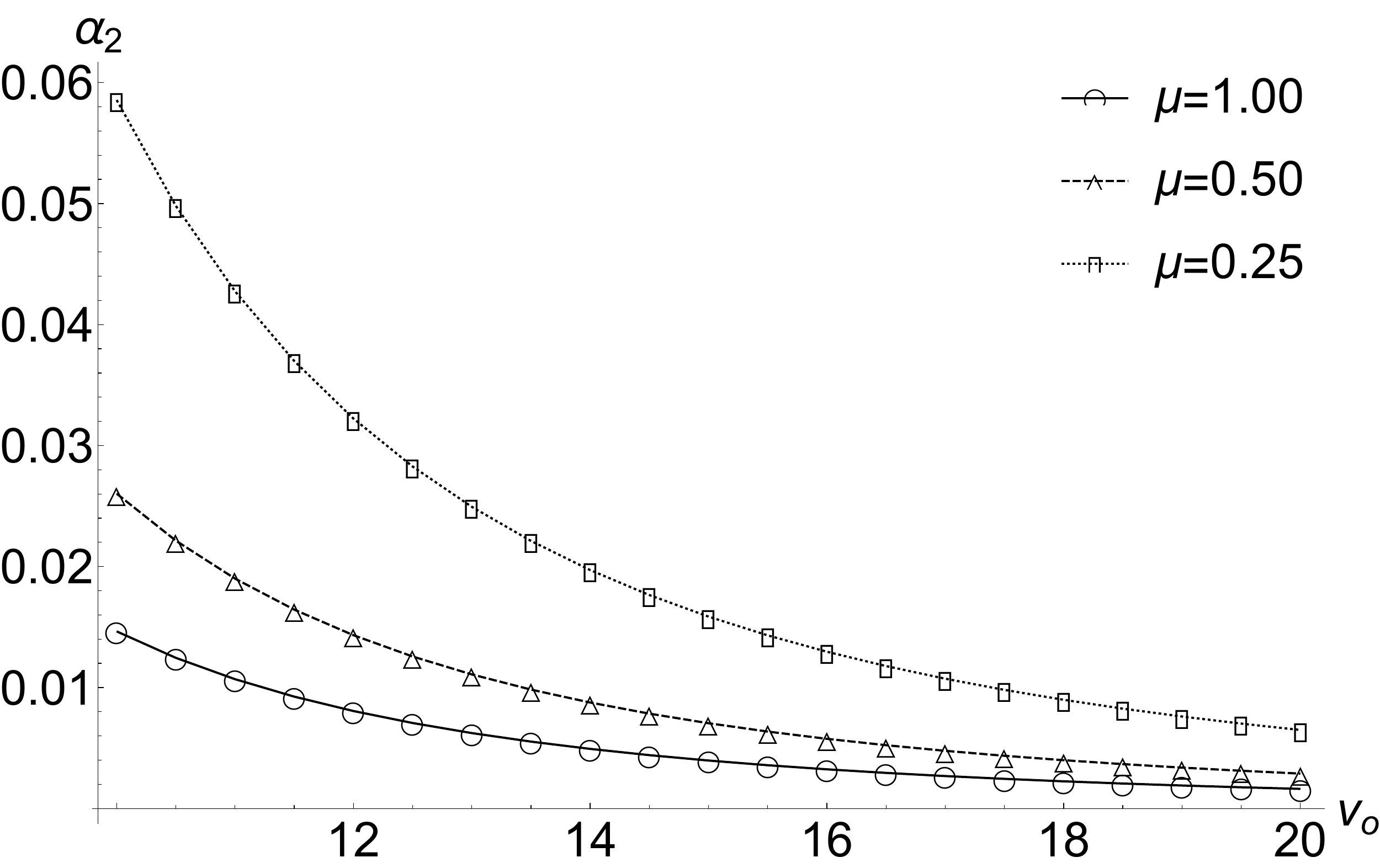}
	\caption{Magnitude of the leading quantum correction term to the classical TOA for a single atom described as a Gaussian wavepacket with increasing initial velocity for the parameters $g=\hbar=1$, initial position $q_o=-5$, arrival point at the origin $q=0$, and $\sigma^2=0.1$.}
	\label{fig:mag_qcorr_plot}
\end{figure}

\section{Quantum correction for Gaussian wavepackets}
\label{sec:quant_corr}
Let us now consider a single particle described by the Gaussian position probability distribution,
\begin{equation}
\varphi(q) = \dfrac{1}{\sqrt{\sigma \sqrt{2 \pi}}} e^{-\frac{1}{4\sigma^2}(q-q_o)^2}.
\label{gaussianwavefunction}
\end{equation}
A closed form solution of the $r^{th}$ order quantum correction term can be obtained by using the definition of the Hermite polynomial
\begin{equation}
H_n(z) = (-1)^n e^{z^2}\frac{d^n}{dz^n}e^{-z^2}
\end{equation}
and using the identity \cite{hermite}
\begin{equation}
\sum_{q=0}^{r}\dfrac{r!}{(r-q)!q!}H_q(x)H_{r-q}(y)=2^{r/2}H_r\left(\dfrac{x+y}{\sqrt{2}}\right).
\end{equation}
Assuming that the spread of the Gaussian wavepacket is sufficiently small, the $r^{th}$ order quantum correction term will have the form      
\begin{align}
\alpha_{r} =& \frac{q_o}{v_o} \left(\dfrac{-2^{5/2}}{4i\sigma \mu v_o}\right)^r \dfrac{\Gamma(\frac{r+1}{2})\Gamma(\frac{r+2}{2})}{\Gamma(\frac{1-r}{2})} \dfrac{1}{r!} \nonumber \\
& \times{_2 F_1}\left(\frac{r+1}{2}, \frac{r+2}{2};2;\dfrac{2 g q_o}{v_o^2}\right)
\label{qcorr}
\end{align}
where ${_2F_1}(a,b;c,z)$ is a specific hypergeometric function.

It can be seen that the $r^{th}$ order quantum correction term is mass dependent, implying a violation of the WEP despite imposing the validity of the first and second statement of the EPCS to be carried over to quantum systems. Since $\Gamma(\frac{1-r}{2})$ is infinite when $r$ is odd, this means that only the terms when $r$ is even will survive and that the quantum correction terms are in even orders of $\hbar$. Up to the leading quantum correction term, the expectation value of the quantum TOA is now 
\begin{align}
\bar{\tau}_{\text{quant}}=&\dfrac{v_o}{g} \left(1 - \sqrt{1-\dfrac{2 g}{v_o^2}q_o} \right) \nonumber \\
& + \frac{q_o}{4 \sigma ^2 \mu ^2 v_o^3}\left(1-\frac{2 g q_o}{v_o^2}\right)^{-3/2}\hbar^{2} + \mathcal{O}(\hbar^4)
\label{expansionconfirm}
\end{align}
where we used the identity ${_2F_1}(a,b;b;z)={_1F_0}(a;;z)=(1-z)^{-a}$. 
The leading quantum correction term $\alpha_2$ is positive, making $\bar{\tau}_{\text{quant}} > \tau_{\text{class}}$. This implies that the particles arrive later than what is expected classically during its first time crossing at the origin.

\begin{figure}[t!]
\begin{subfigure}
\centering
\includegraphics[width=0.4\textwidth]{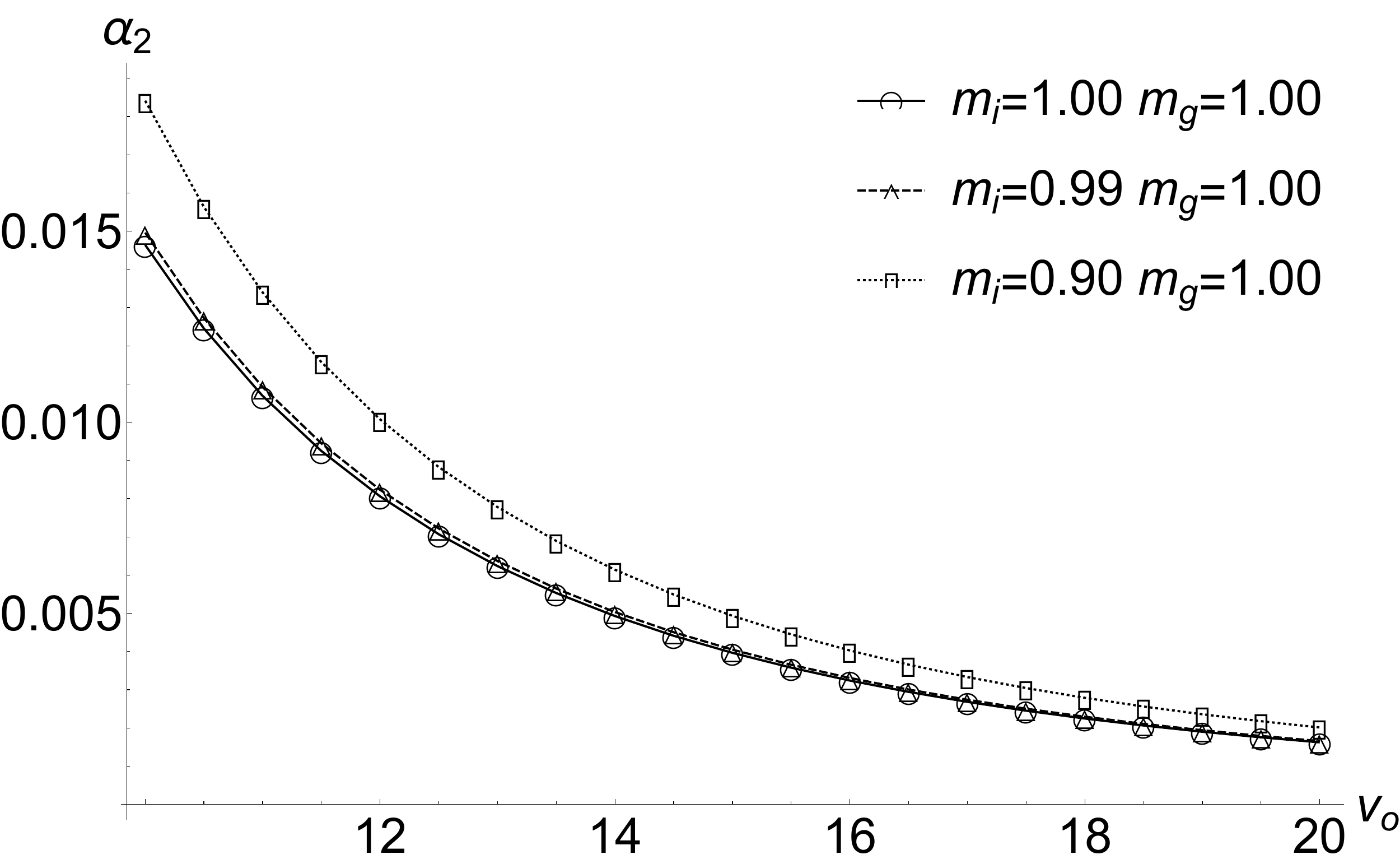}
\label{fig:mi_less_mg}
\end{subfigure}
\begin{subfigure}
\centering
\includegraphics[width=0.4\textwidth]{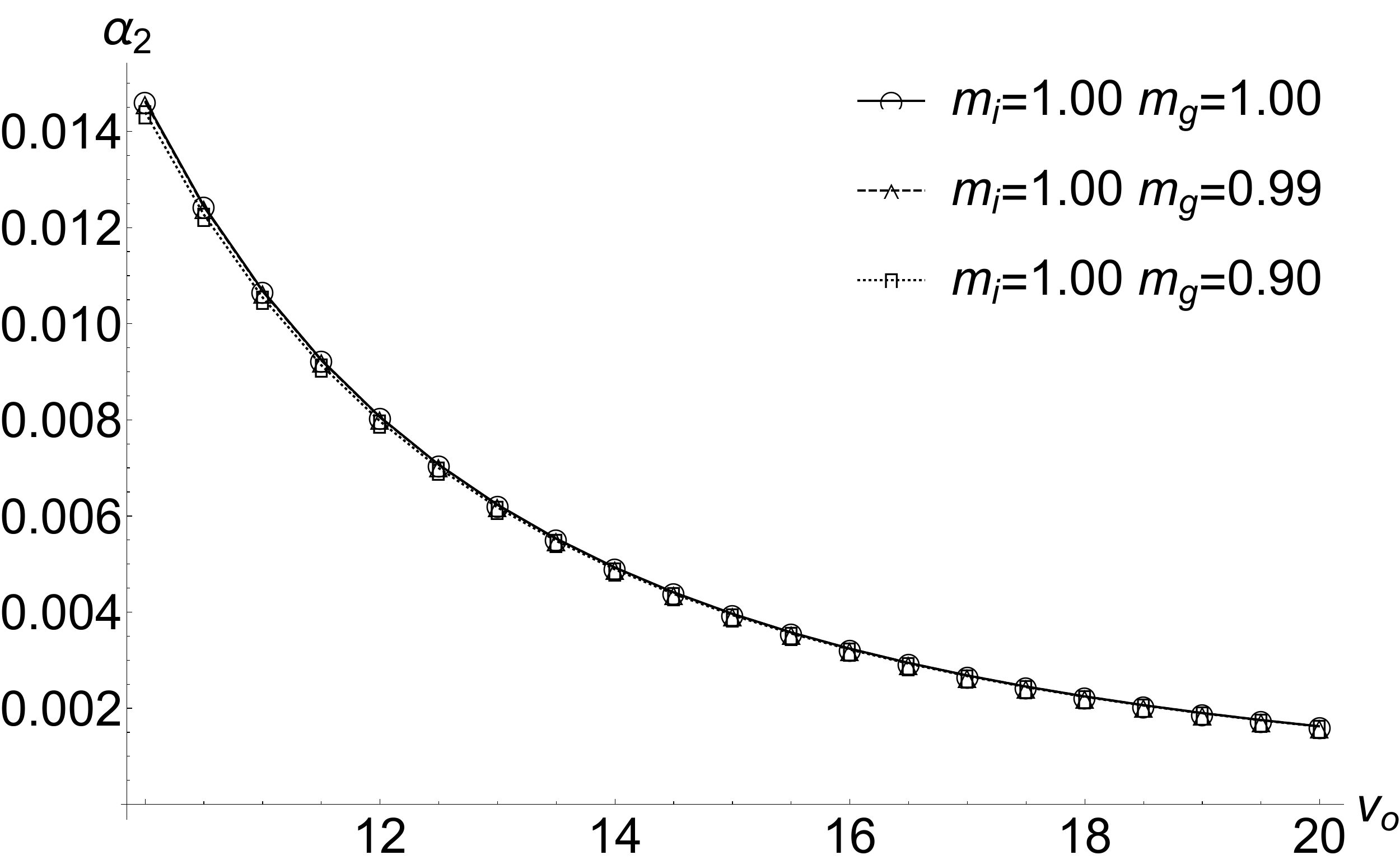}
\label{mi_greater_mg}
\end{subfigure}
\begin{subfigure}
\centering
\includegraphics[width=0.4\textwidth]{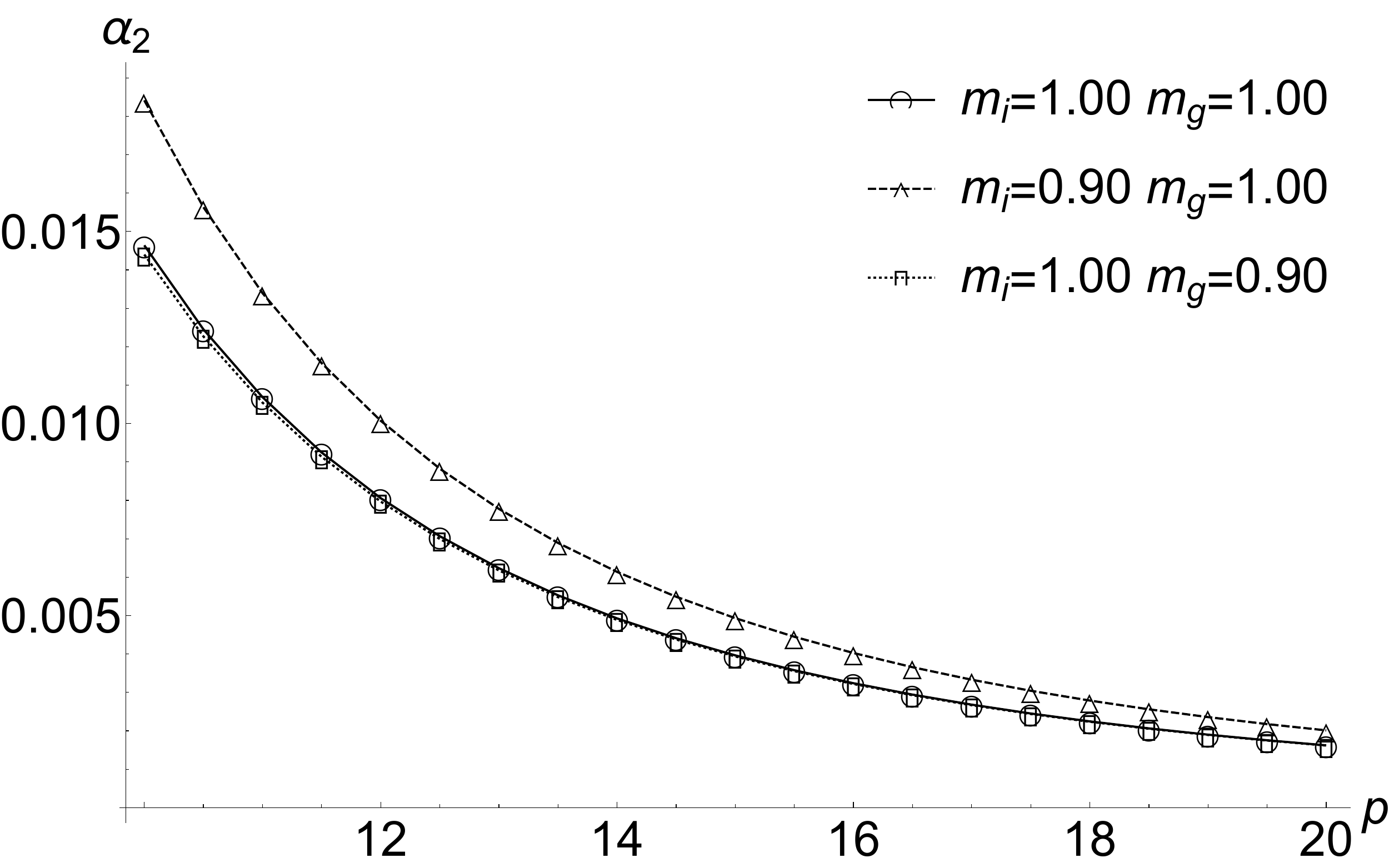}
\label{fig:mi_nequal_mg}
\end{subfigure}
\caption{Magnitude of the leading quantum correction term to the classical TOA for a single atom described as a Gaussian wavepacket with increasing initial velocity for the parameters $g=\hbar=1$, initial position $q_o=-5$, arrival point at the origin $q=0$, and $\sigma^2=0.1$ for $m_i \neq m_g$.}
\label{fig:leadingquantumcorrection}
\end{figure}

\begin{figure*}[t!]
\begin{subfigure}
\centering
\includegraphics[width=0.4\textwidth]{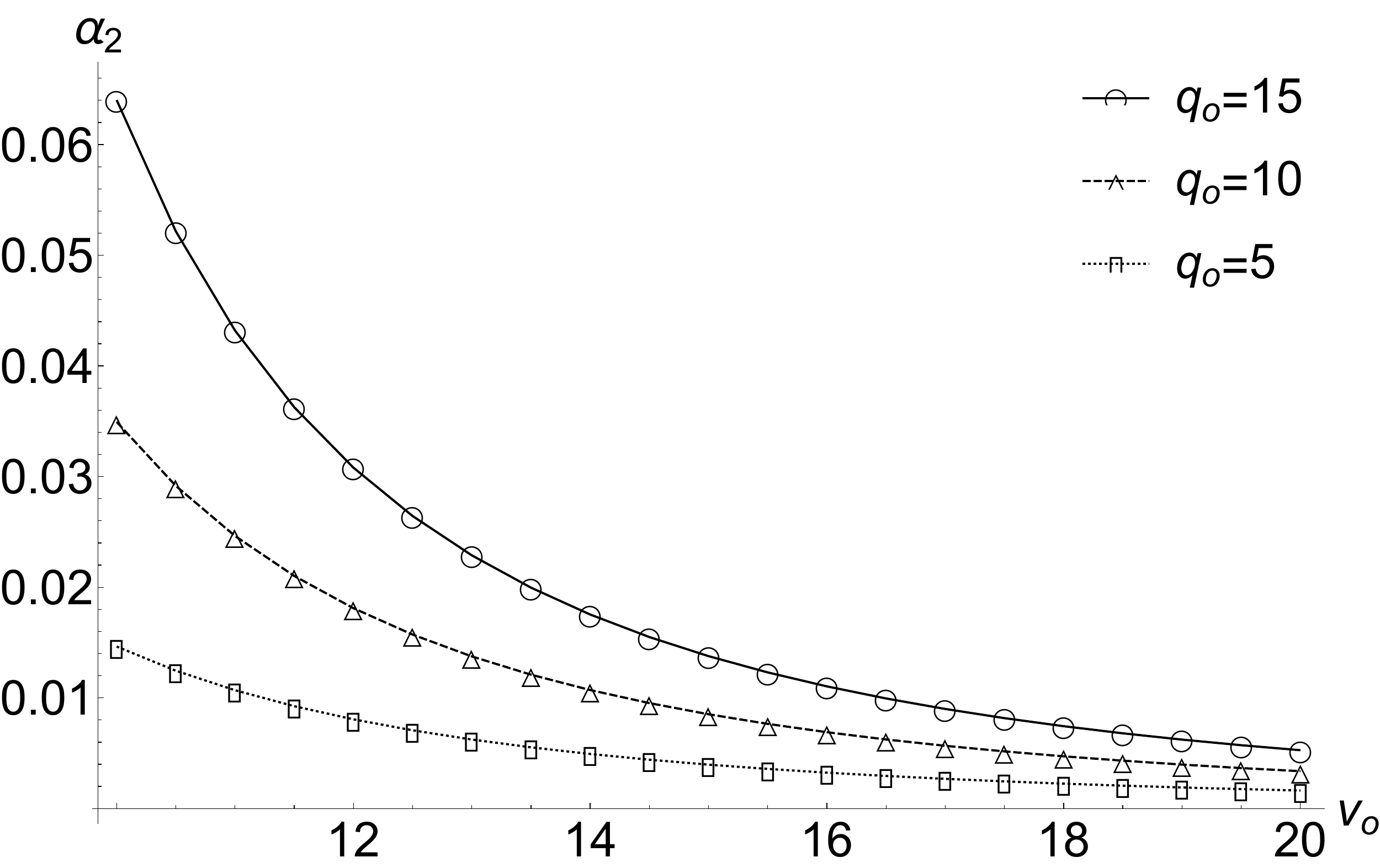}
\label{fig:diffqo}
\end{subfigure}
\begin{subfigure}
\centering
\includegraphics[width=0.4\textwidth]{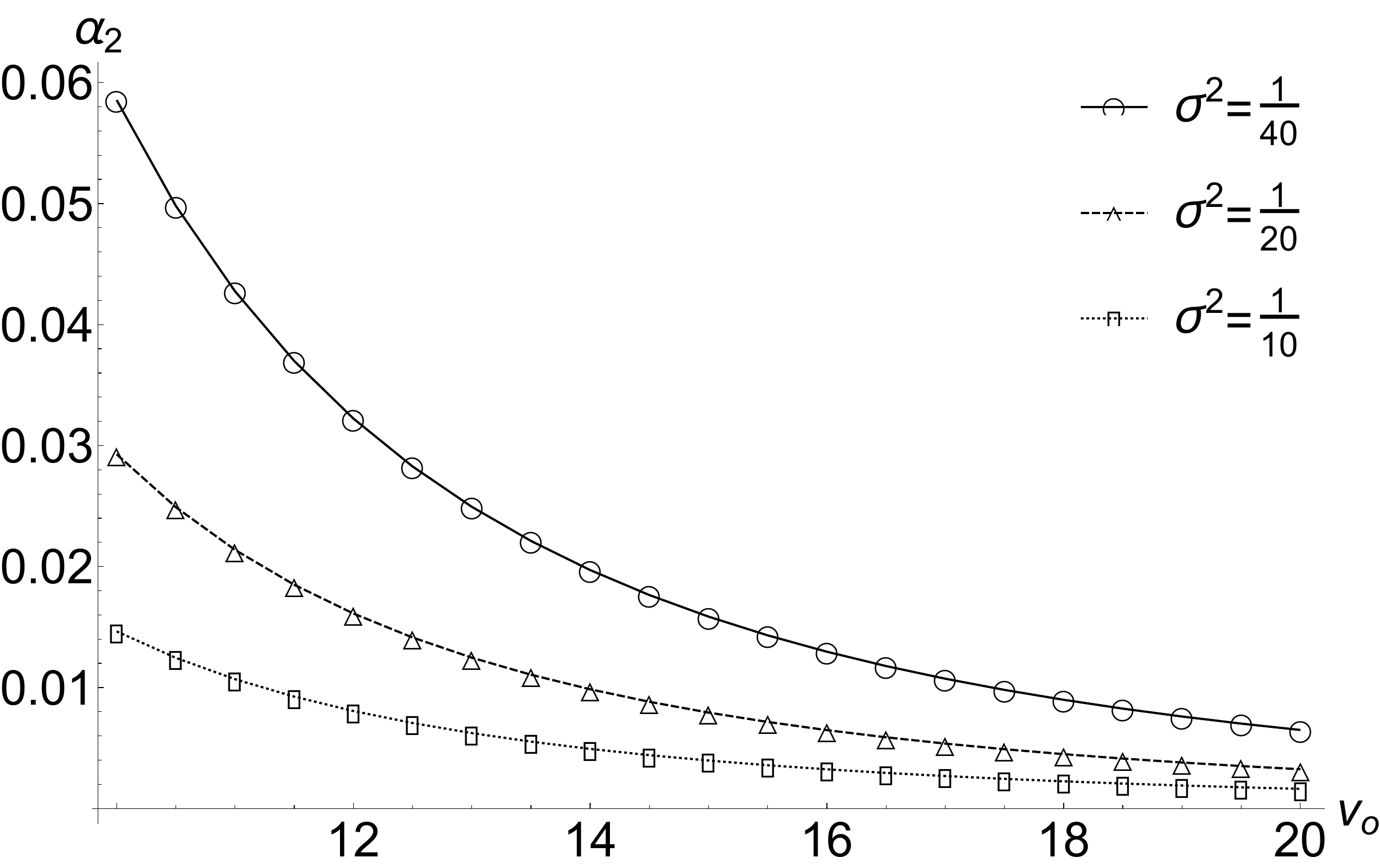}
\label{fig:diffsigma}
\end{subfigure}
\caption{Magnitude of the leading quantum correction term to the classical TOA for a single atom described as a Gaussian wavepacket with increasing initial velocity for the parameters $\mu=g=\hbar=1$, arrival point at the origin $q=0$, and $\sigma^2=0.1$ for $\sigma^2=0.1$ with varying initial position (top), and for $q_o=-5$ with varying spread of the wavepacket.}
\label{fig:sigmaandqo}
\end{figure*}

We now address the issue of missing terms raised above by numerically confirming the accuracy of the semiclassical expansion Eq. \eqref{expansionconfirm} against the exact expression given by Eq. \eqref{toaexpec}. We first consider the case when the incident wavepacket has a spread or support that is sufficiently small such that $2g q_o /v_o^2 <1$. Choosing the parameters $\mu=g=\hbar=1$ with initial position $q_o=-5$, inital velocity $v_o=30$, and $\sigma^2=0.1$. The exact expression given by Eq. \eqref{toaexpec} yields $\tau=0.166663$. On the other hand, the semiclassical expansion, up to the leading quantum correction, yields $\tau=0.166662$, where the classical value is $\tau_o=0.166206$, and the leading quantum correction is $0.000455$. Thus, the accuracy of the the semiclassical expansion confirms that the interchange of the order of integration and summation has lead to a negligible contribution from the missing terms under the  condition that $2g q_o /v_o^2 < 1$ \cite{wongBook,wongPAMS1980,galaponPRSA2017}.

However, the semiclassical expansion Eq. \eqref{expansionconfirm} fails when $2g q_o /v_o^2 > 1$. For example, choosing the parameters $\mu=g=\hbar=1$ with initial mean position $q_o=-5$, initial mean velocity $v_o=2$, and $\sigma^2=0.1$, the leading term in the semiclassical expansion Eq. \eqref{expansionconfirm} yields the complex value $\tau=2.0-1.598972i$. It is not difficult to show that the real part of the complex $\tau$ is equal to the (classical) time the particle will arrive at the (classical) turning point, which, for the given parameters, is below the intended arrival point which is the origin. This can be readily verified for the parameters in consideration. Now the exact expected time of arrival computed 
from expression Eq. \eqref{toaexpec} is $\tau=3.918569$; this is greater than the classical time of arrival at the turning point which is equal to $2.0$. The extra time arises from the additional time the particle has to take to tunnel to the origin from the turning point. Thus our semi-classical expansion is accurate when tunneling effect is negligible.  

Now the magnitude of the leading quantum correction term is shown in Fig. \ref{fig:mag_qcorr_plot}. It can be seen that massive particles have a smaller quantum correction to its classical TOA, and as the initial velocity increases, the quantum correction also becomes very small. This indicates that in the limit of large momentum, the leading quantum correction term becomes negligible. Therefore, we recover the classical behavior of TOA for a particle fired upward. Suppose we consider a ${^{133}}Cs$ atom (which is commonly used in atomic fountain clocks) that is fired upward with an initial velocity of $v_o=10 m/s$, inital position $q_o=-1 m$, and $\sigma=1 mm$ for the parameters $\hbar=1.05\times 10^{-34}$ and $g=9.8m/s^2$. This gives a leading quantum correction correction to the classical time of arrival which is equal to $7.46\times 10^{-17} s$. 

A mass dependent quantum correction to the classical time of arrival of a particle in free-fall has also been shown in Refs. \cite{mousavi2015,chowdhury2012,viola1997,davies2004a,ali2006}. Nonetheless, the weak equivalence principle has always been recovered either in the limit of large mass or momentum. The quantum correction terms physically arise from the accumulated quantum effects of the particle as it is fired upward, e.g. spreading of the wavepacket as it evolves through time, quantized vertical motion of the particle, and backscattering before the particle reaches the classical turning point. Compared to the other methods used in Refs. \cite{mousavi2015,chowdhury2012,viola1997,davies2004a,ali2006} which used the initial wavefunction and the time-evolved wavefunction to calculate the time of arrival of the particles under the influence of a gravitational potential, our method only uses the initial wavefunction to calculate the time of arrival. The reason for this is that, in our treatment, time of arrival is a dynamical observable which is represented by a Hermitian operator from which the expected time of arrival is obtained from the expectation value of the time of arrival operator. 

Using Eq. \eqref{qcorr}, we can also investigate the effects of the quantum correction terms if we did not assume the inertial and gravitational masses to be equal. To do this, we just perform a change of variables $\mu\rightarrow m_i$ and $g\rightarrow m_g g / m_i$. Fig. \ref{fig:leadingquantumcorrection} shows that the ratio $m_i/m_g$ has a role on the magnitude of the leading quantum correction. That is, if the ratio $m_i/m_g>1$ there is no significant effect on the leading quantum correction term but when but there is a significant effect when $m_i/m_g<1$. However, this effect becomes negligible as $m_i/m_g \rightarrow 1$. 

The effect of the preparation of the initial state on the leading quantum correction, such as the spread of the wavepacket $\sigma$ and initial position $q_o$, can also be investigated using Eq. \eqref{qcorr}. Fig. \ref{fig:sigmaandqo} (top) shows that the leading quantum correction becomes larger as the arrival point becomes closer to the classical turning point. This result is also consistent with that of Davies in Ref. \cite{davies2004a}. Meanwhile, Fig. \ref{fig:sigmaandqo} (bottom) shows that as the spread of the initial wavepacket becomes larger the leading quantum correction term becomes smaller.

\begin{figure}[t!]
\begin{subfigure}
\centering
\includegraphics[width=0.4\textwidth]{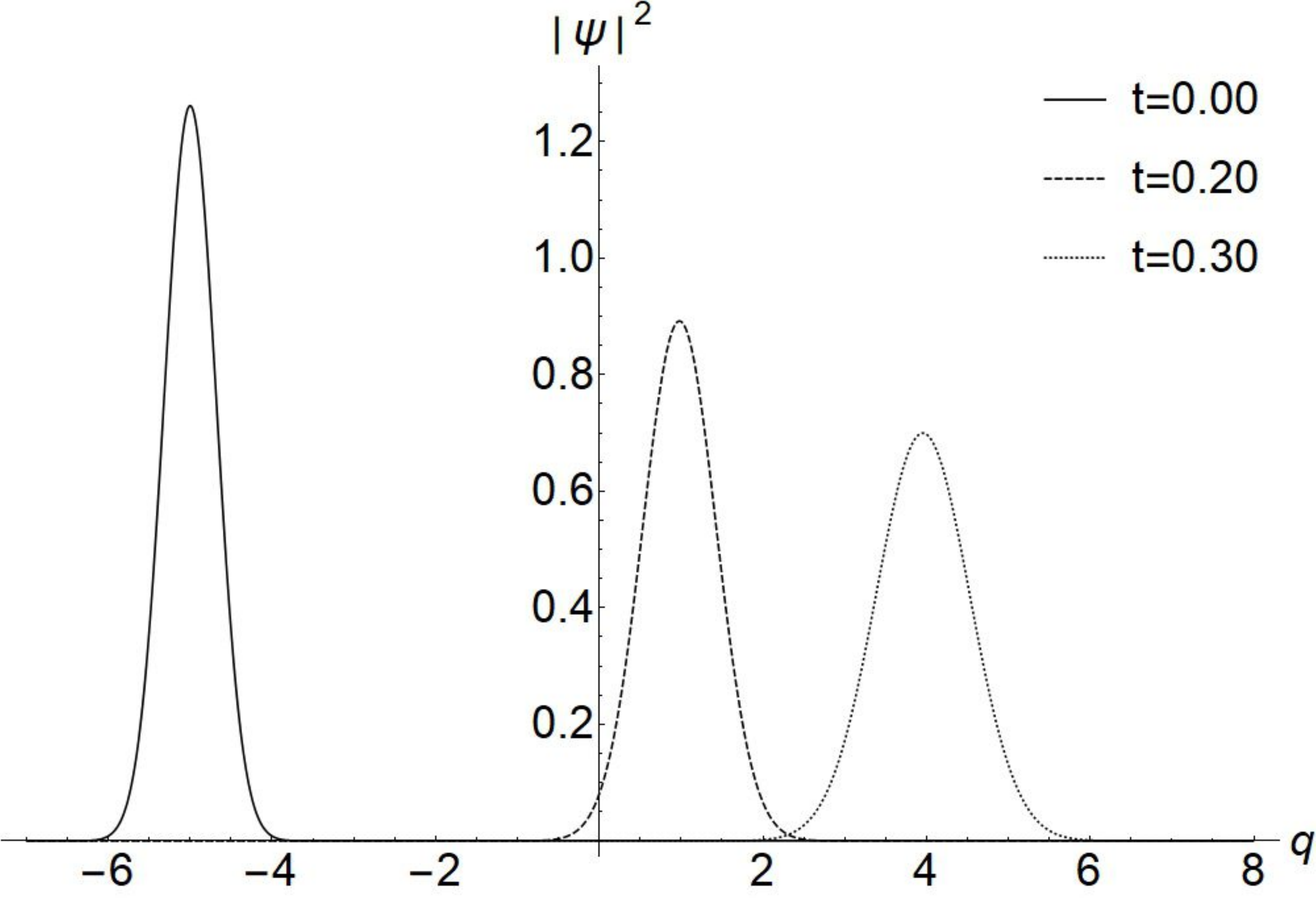}
\label{fig:pos_prob}
\end{subfigure}
\begin{subfigure}
\centering
\includegraphics[width=0.4\textwidth]{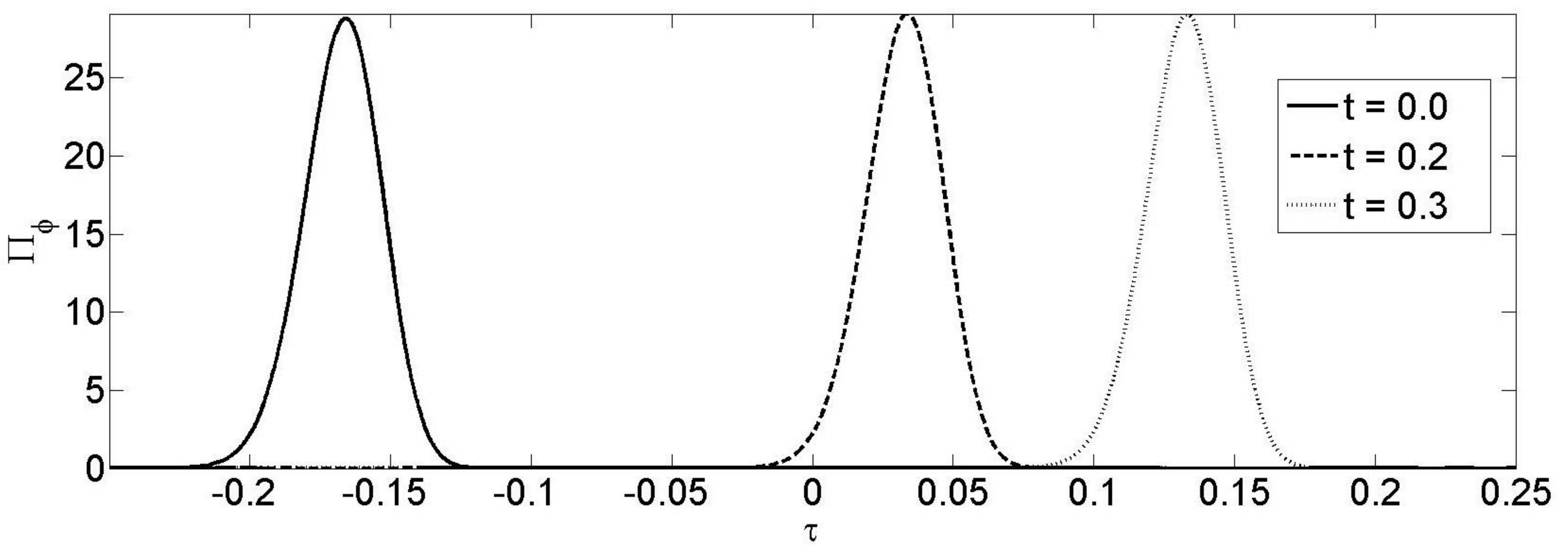}
\label{fig:toa_dist}
\end{subfigure}
\caption{The time evolved position density distribution of a single atom described as a Gaussian wavepacket with initial velocity $v_o=30$, initial position $q_o=-5$, and $\sigma^2=0.1$ for the parameters $\mu=g=\hbar=1$ (top) with its corresponding time of arrival distribution at the origin (bottom).}
\label{fig:covariance}
\end{figure}

\section{Time of arrival distribution for single particles}
\label{sec:toa_dist}

In quantum mechanics, we do not expect that an ensemble of identical particles prepared in the same initial state will arrive at the origin at the same time but rather, we get a TOA distribution at the origin which should peak at the expectation value of the TOA operator. This does not necessarily imply violation of the weak equivalence principle but may well be a consequence of the probabilistic nature of quantum mechanics. If we assume that the second statement of the AHWEP is indeed true, then the TOA distribution of two different particles should be identical as long as the initial group velocities are equal. Consequently, nonidentical TOA distributions can be used to distinguish the particles form each other, which implies a violation of the WEP. 

\begin{figure}[t]
	\centering
	\includegraphics[width = 0.5\textwidth]{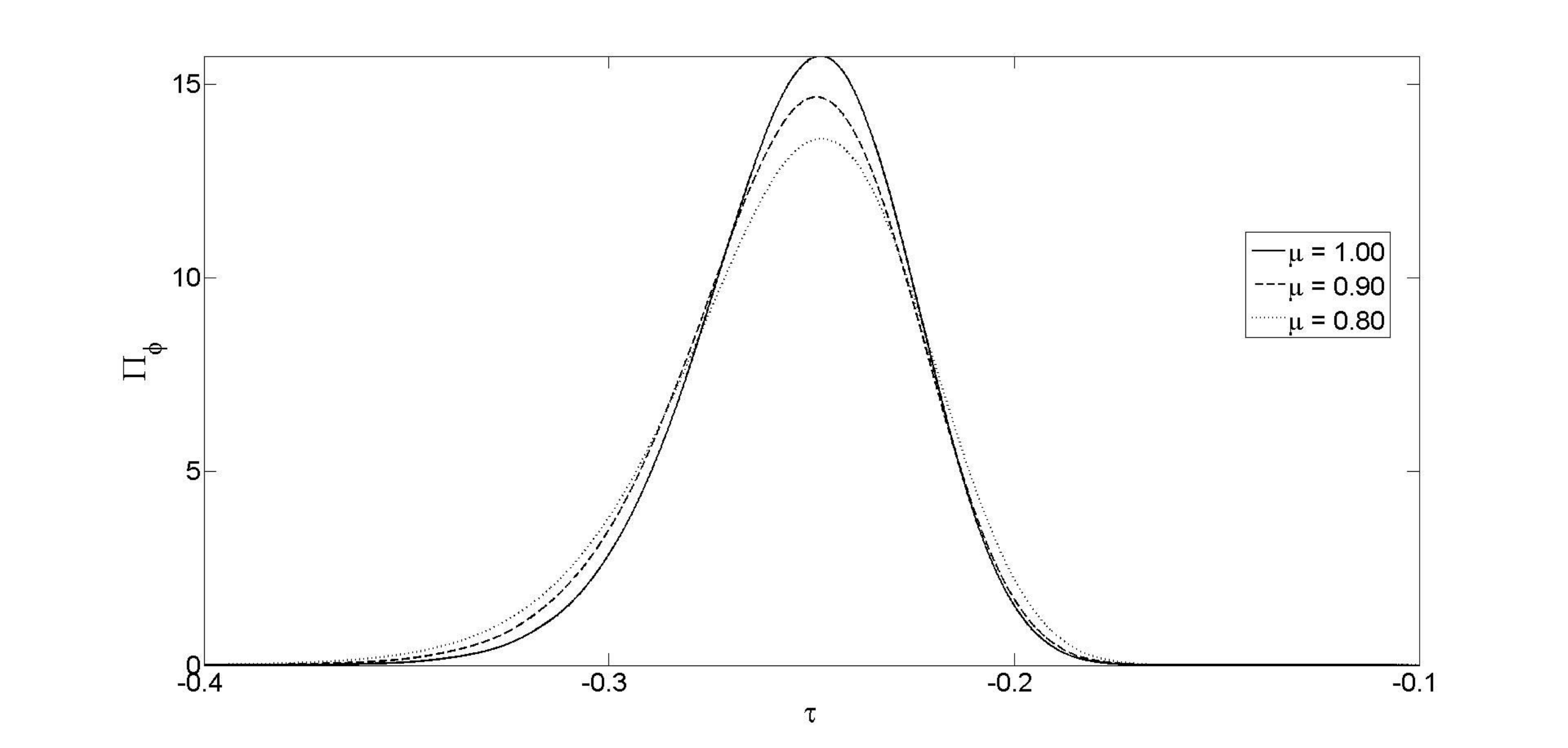}
	\caption{Time of arrival distribution of a single atom described as a Gaussian wavepacket for the parameters $g=\hbar=1$ with initial initial velocity $v_o=20$, initial position $q_o=-5$, arrival point at the origin $q=0$, and $\sigma^2=0.1$}
	\label{fig:dist_mieqmg}
\end{figure}

To construct the TOA distribution for the single particles, we start by defining the probability that a particle in state $\phi$ will arrive at the origin, at a time $t$ before $\tau$ as
\begin{align}
\expec{\phi}{\opr{\Pi}}{\phi} =& \int_{-\infty}^{\tau}\dirac{\phi}{t}\dirac{t}{\phi} dt \nonumber \\
=&\int_{-\infty}^{\tau}\int_{-\infty}^{\infty}\int_{\infty}^{\infty} \dirac{\phi}{q'}\dirac{q'}{t}\dirac{t}{q} \nonumber \\
&\times \dirac{q}{\phi} dt dq' dq
\label{probbeftau}
\end{align}
where $| t \rangle$ is an eigenket of the TOA operator, and $\opr{\Pi}$ is the positive operator valued measure corresponding to the TOA distribution. The TOA distribution at the origin can thus be constructed by differentiating Eq. \eqref{probbeftau} with respect to $\tau$, explicitly we get
\begin{equation}
\Pi_{\phi_o}(q,\tau) = \dfrac{d}{d\tau}\expec{\phi}{\opr{\Pi}}{\phi} = \abs{\int_{-\infty}^{\infty}\phi^{*}(q)\psi(\tau,q)dq}^2 
\label{toadistribution}
\end{equation}
where $\psi(\tau,q)$ is an eigenfunction of the TOA operator. The construction of the TOA-distribution by quadrature is discussed in detail in Appendix-B.

\begin{figure}[t!]
\begin{subfigure}
\centering
\includegraphics[width=0.5\textwidth]{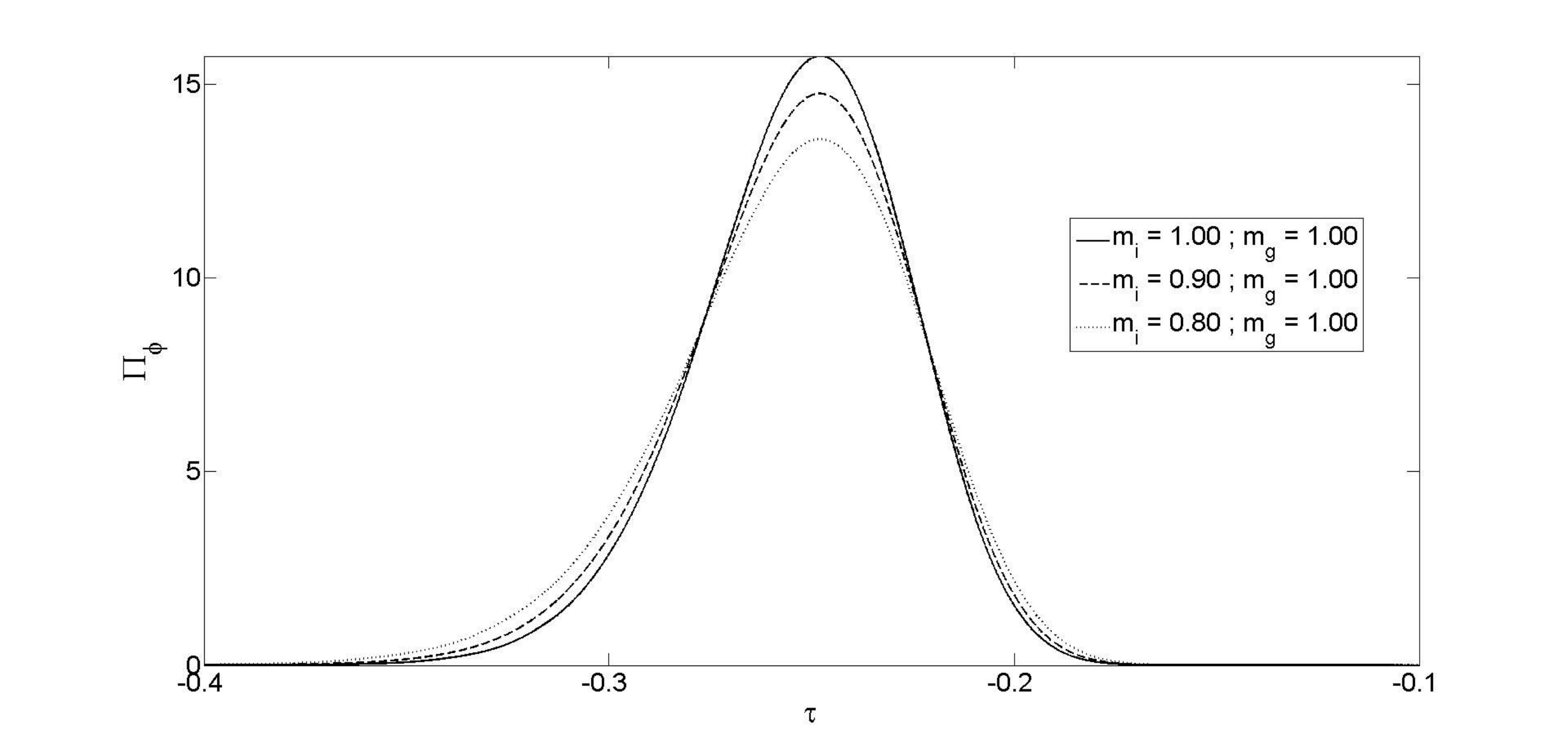}
\label{fig:dist_milemg}
\end{subfigure}
\begin{subfigure}
\centering
\includegraphics[width=0.5\textwidth]{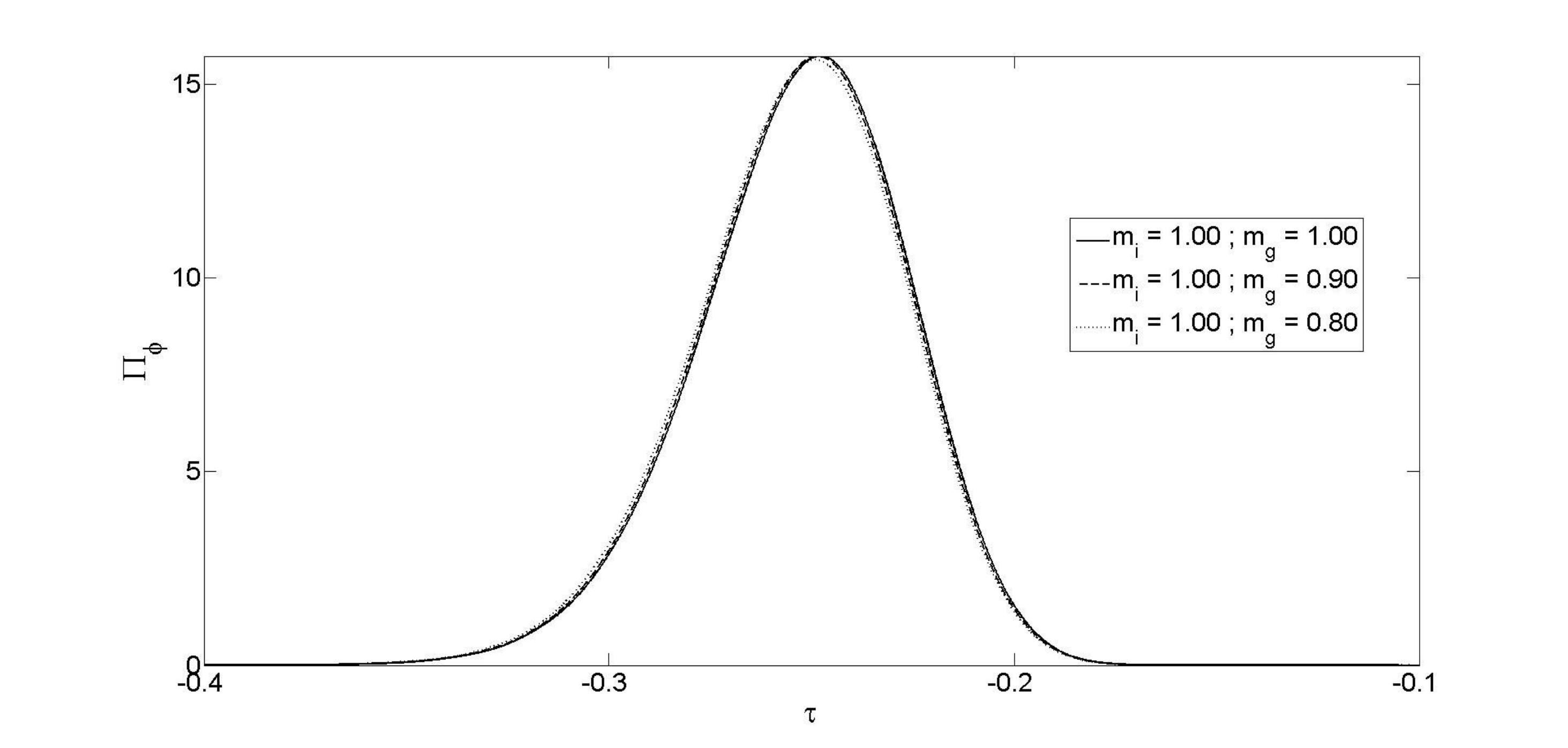}
\label{fig:dist_migrmg}
\end{subfigure}
\begin{subfigure}
\centering
\includegraphics[width=0.5\textwidth]{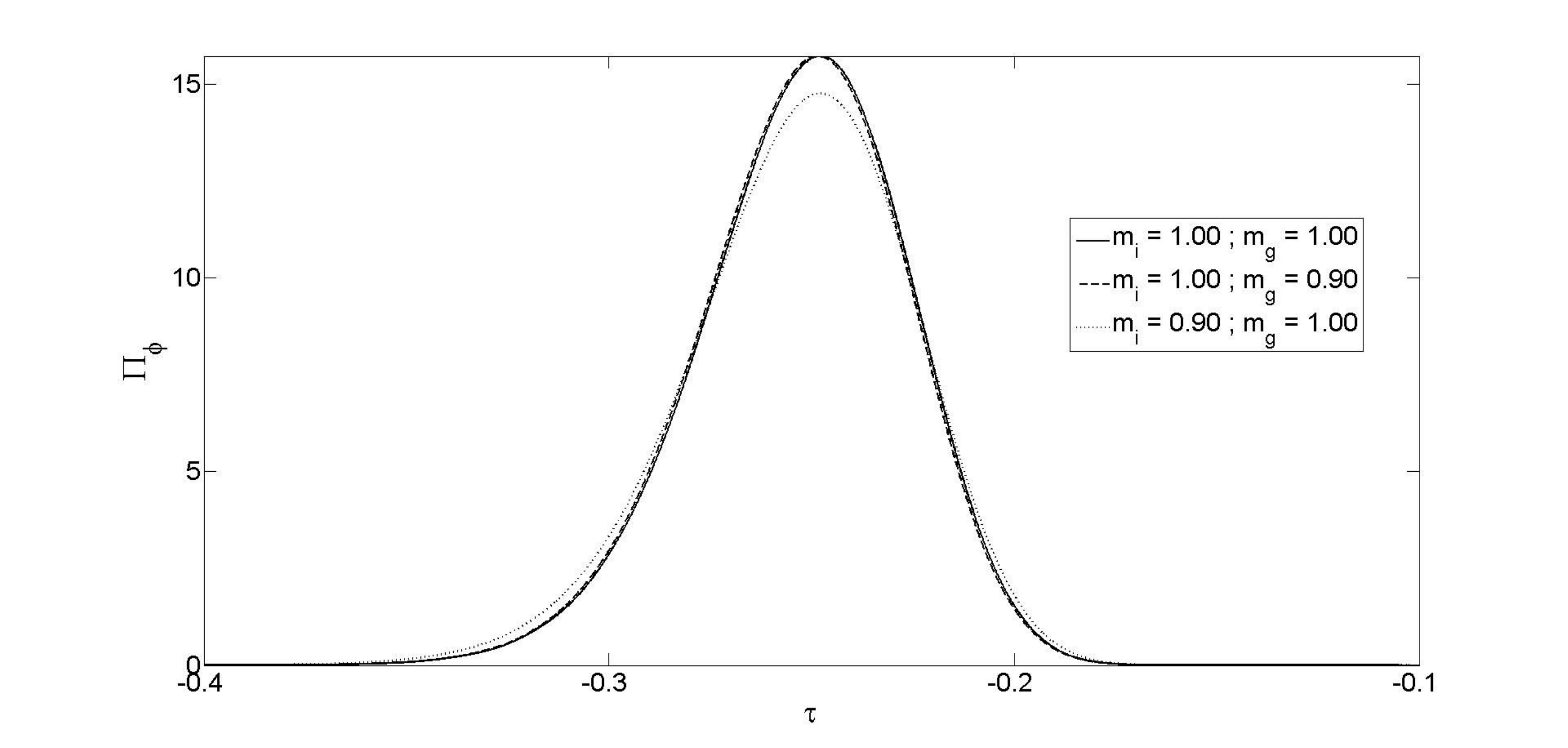}
\label{fig:dist_minemg}
\end{subfigure}
\caption{Time of arrival distribution of a single atom described as a Gaussian wavepacket for the parameters $g=\hbar=1$ with initial velocity $v_o=20$, initial position $q_o=-5$, arrival point at the origin $q=0$,  and $\sigma^2=0.1$ for $m_i \neq m_g$.}
\label{fig:toadist_overlay}
\end{figure} 

\begin{figure}[t!]
\begin{subfigure}
\centering
\includegraphics[width=0.5\textwidth]{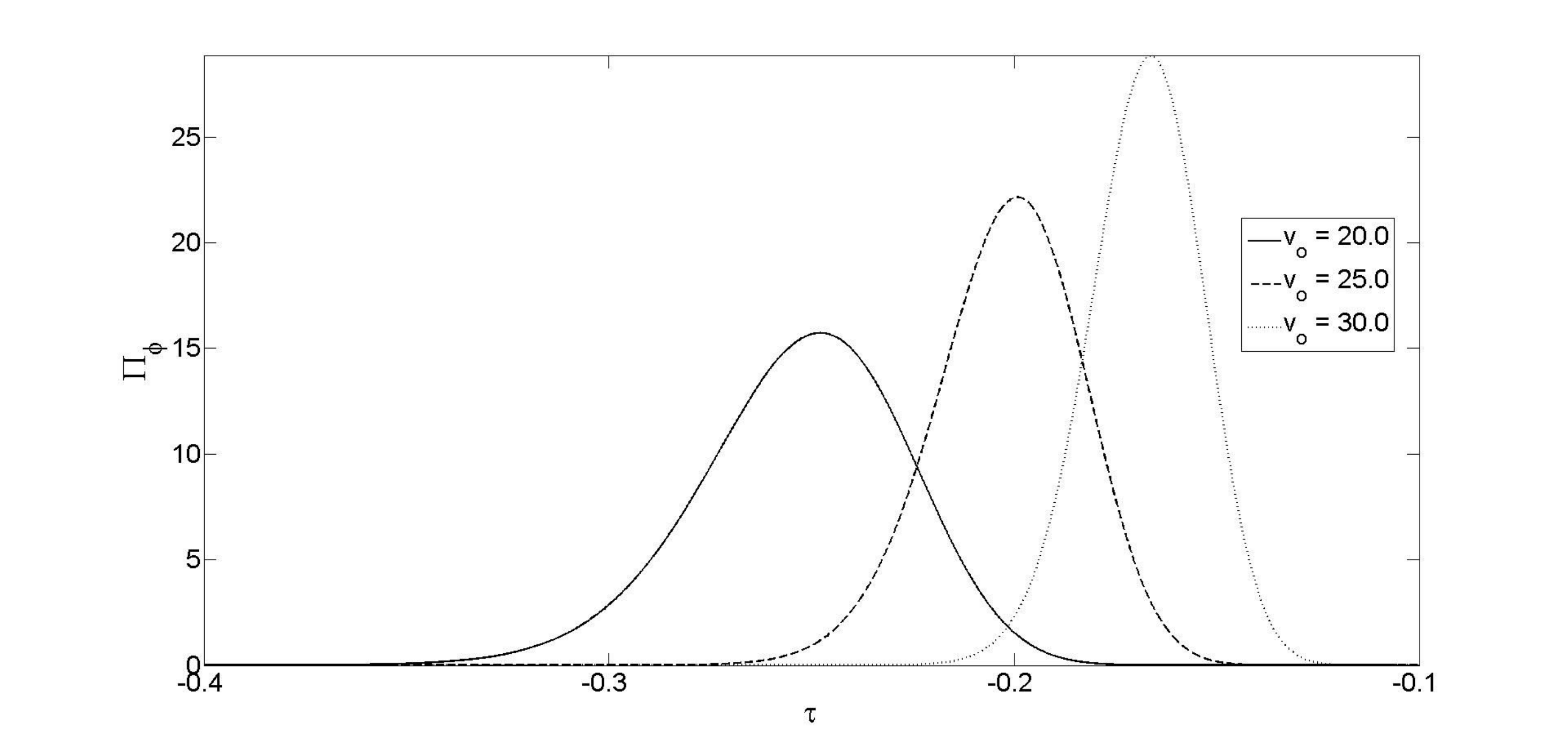}
\label{fig:dist_diff_po}
\end{subfigure}
\begin{subfigure}
\centering
\includegraphics[width=0.5\textwidth]{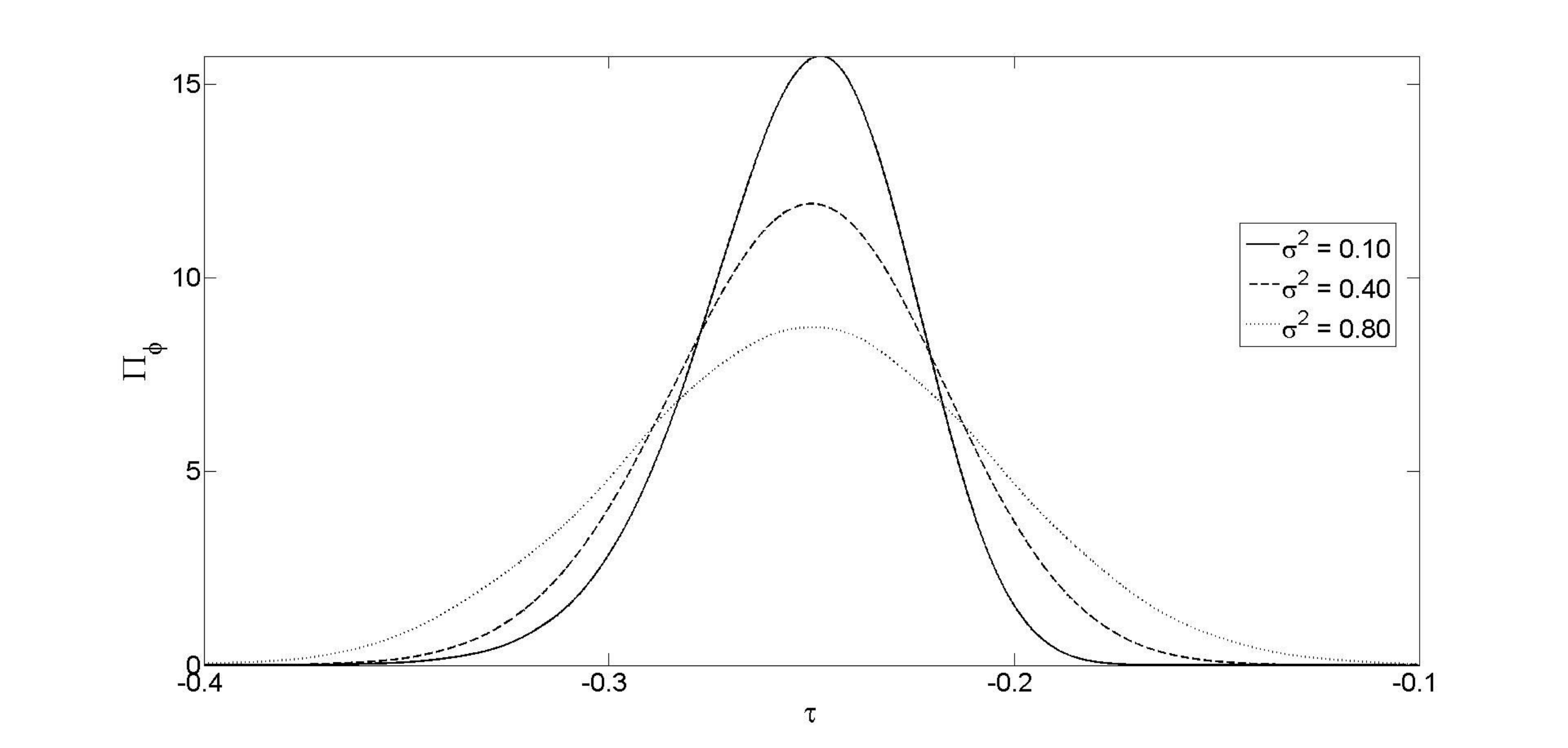}
\label{fig:dist_diff_sigma2}
\end{subfigure}
\begin{subfigure}
\centering
\includegraphics[width=0.5\textwidth]{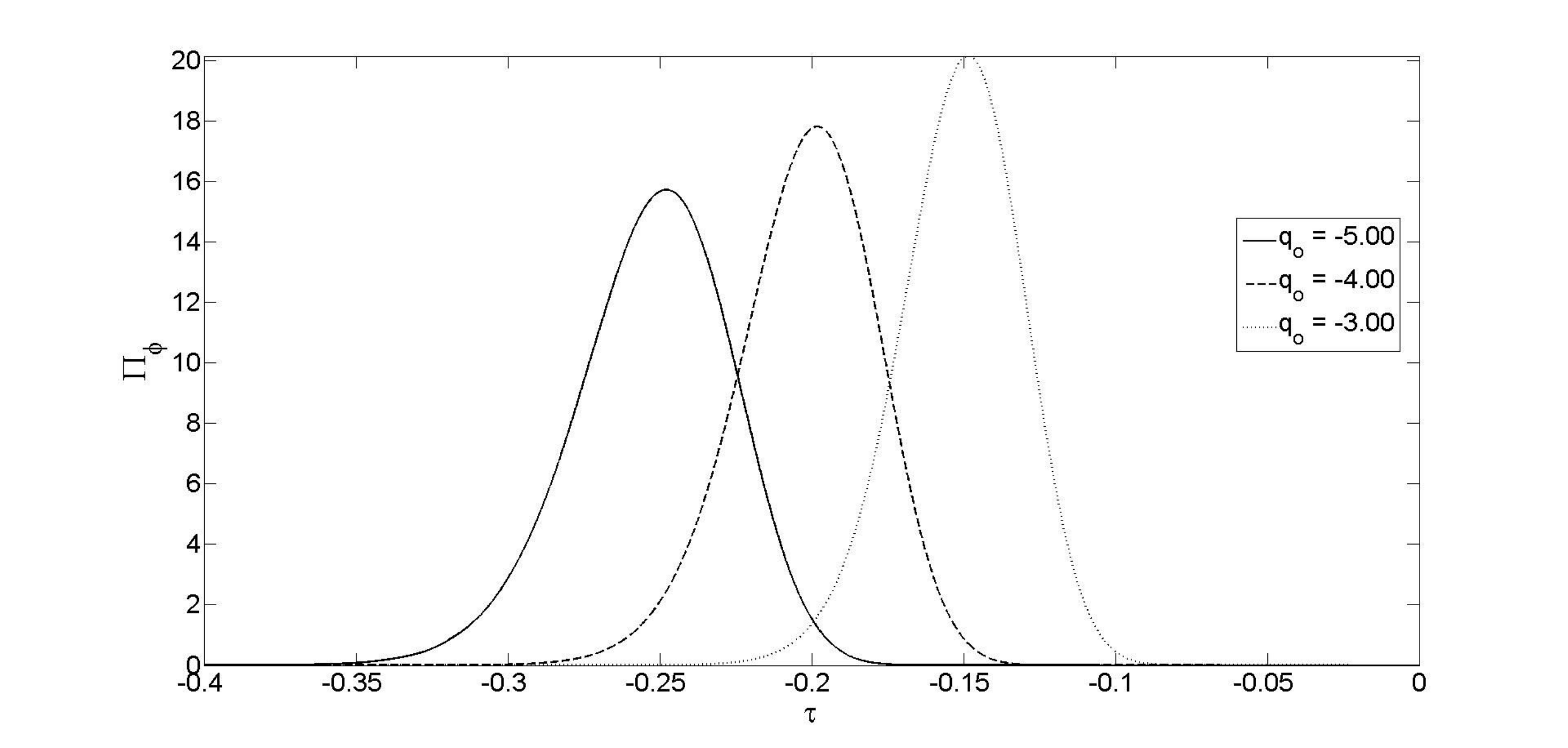}
\label{fig:dist_diff_qo}
\end{subfigure}
\caption{Time of arrival distribution of a single atom described as a Gaussian wavepacket for the parameters $\mu=g=\hbar=1$ with $\sigma^2=0.1$ and $q_o=-5$ (top), $v_o=20$ and $q_o=-5$ (middle), and $v_0=20$ and $\sigma^2=0.1$ (bottom).}
\label{fig:toadist_other}
\end{figure}

The conjugacy of the TOA operator with the system Hamiltonian implies covariance, i.e. the Hamiltonian and the TOA operator should be generators of translation of each other. Particularly, if $\Pi_{\phi_o}(q,\tau)$ is the the TOA distribution of a given initial state $\phi_o=\phi(q,t=0)$, then the TOA distribution for the time-evolved state $\phi(q,t)=\opr{U}_t \phi(q,t=0)$ is 
\begin{equation}
\Pi_{\phi}(q,\tau-t) = \abs{\int_{-\infty}^{\infty}\phi^{*}(q,t)\psi(\tau,q)dq}^2
\label{toadistributiontevo}
\end{equation}
where $\opr{U}_t$ is the time evolution operator. This implies that $\Pi_{\phi}(q,\tau-t)$ is just a time translation of $\Pi_{\phi}(q,\tau)$. We show that the constructed TOA operator is covariant under time translation by evolving the initial wavefunction described by Eq. \eqref{gaussianwavefunction}, using the well-known linear potential propagator
\begin{align}
K(q,t;q',0)=&\sqrt{\dfrac{\mu}{2\pi i \hbar t}}\exp\left[i\dfrac{\mu(q-q')^2}{2 t \hbar}-i\dfrac{\mu g (q+q')t}{2\hbar}\right]\nonumber \\
&\times \exp\left[-i\dfrac{\mu g^2 t^3}{24 \hbar}\right],
\end{align}
which yields the time-evolved wavefunction as
\begin{align}
\phi(q,t)=& \dfrac{1}{\sqrt{s_t \sqrt{2\pi}}}\exp\left[-\dfrac{(q-q_o-v_ot+\frac{1}{2}gt^2)^2}{4 s_t \sigma}\right] \nonumber \\
&\times \exp\left[i\dfrac{\mu v_o}{\hbar}q_o\right]\exp\left[-i\dfrac{\mu q_o}{\hbar}gt\right]\exp\left[-i\dfrac{1}{6}\dfrac{\mu}{\hbar}g^2t^3\right] \nonumber \\
& \times \exp\left[i\dfrac{\mu}{\hbar}\left(v_o-gt\right)\left(q-q_o-\frac{1}{2}v_o t\right)\right] 
\end{align}
where $s_t=\sigma(1+i\hbar t/ 2 \mu \sigma^2)$. The position density distribution and time of arrival distribution for the first time crossing at the origin are both plotted in Fig. \ref{fig:covariance}. It can be seen that the particle follows a classical trajectory as expected, and it can also be seen that the TOA distribution of the time-evolved wavefunction are just time translations of each other (See Appendix-B for a discussion in solving the eigenvalue problem for the time of arrival operator $\opr{T}$.).

With covariance established, we now exhibit the time of arrival distribution for the first time crossing of particle fired upwards. Now, if the weak equivalence principle does hold for quantum systems then the TOA distribution for different particles should be identical regardless of mass and composition as long as the particles have the same initial velocity. However, as can bee seen in Fig. \ref{fig:dist_mieqmg}, the time of arrival distribution for different masses with the same initial group velocity are distinguishable from each other. Furthermore,  the peaks of the three TOA distributions do not coincide even though they have the same initial group velocity. The shift in the peaks is attributed to the quantum corrections to the classical TOA. These then imply that particles can be differentiated from each other based on their different time of arrival distributions, which means a violation of the weak equivalence principle. 

We can also investigate the effects on the TOA distribution if we did not assume the equivalence of the inertial and gravitational masses. It can be seen from Fig. \ref{fig:toadist_overlay} (top) that if the ratio $m_i/m_g<1$, then the time of arrival distribution will have a noticeable change. Meanwhile, Fig. \ref{fig:toadist_overlay} (middle) shows that there is a small change in the time of arrival distribution when $m_i/m_g>1$. The effects of the ratio $m_i/m_g$ on the time of arrival distribution is also consistent with its effect on the leading quantum correction term in Sec. \ref{sec:toa_expec}. 

The effect of the preparation of the initial state can also be investigated as shown in Fig. \ref{fig:toadist_other}. The time of arrival distribution becomes sharper as the momentum is increased as shown in Fig. \ref{fig:toadist_other} (top) which is expected since the particle starts to behave classically for higher energies. When the variance in the position of the initial wavefunction increases, the variance of the time of arrival distribution also increases as seen in Fig. \ref{fig:toadist_other} (middle). This then means that the spread of the TOA distribution is larger because the wavepacket becomes more spread out as it reaches the arrival point. Lastly, changing the initial position of the particle causes a shift in the TOA distribution as seen in Fig. \ref{fig:toadist_other} (bottom) while at the same time the TOA distribution becomes sharper as the starting position is closer to the arrival point. This change in the sharpness of the TOA distribution is consistent with that of Fig. \ref{fig:toadist_other} (middle). When the initial position is near the arrival point then it will take less time to reach the arrival point. Since it takes less time to reach the arrival point, the spread of the wavepacket as it reaches the arrival point is smaller than the case when the initial position is far.

\section{Summary and conclusion}
\label{sec:conc}

We addressed the compatibility of the WEP and quantum mechanics by studying the motion of a non-relativistic and structureless quantum particle projected upward in a uniform gravitational field within the context of quantum time of arrival problem by the agency of time of arrival operators. The appropriate time of arrival operator for the projectile was constructed under the constraints of the equivalence of the inertial and gravitational masses, and the  equivalence  of  a  state  of  rest  in  a  homogeneous  gravitational  field  and  the  state  of  uniform acceleration in the absence of gravity. This was accomplished by Weyl-quantization of the mass-independent classical expression for the classical of arrival at the origin using Weyl quantization. 

The mass dependence of the motion of the quantum projectile was investigated by looking at the expectation value of the TOA-operator and the time of arrival distribution for a given initial state. It was found that the expected time of arrival is equal to the classical time of arrival plus mass dependent quantum correction terms in orders of $\hbar$. The magnitude of the correction terms becomes negligible either in the limit of large mass or velocity, so that the WEP is recovered in the limit of 
large incident momentum. Moreover, it was found that the time of arrival distribution depends on the mass of the projectile, specifically, 
massive particles have a sharper TOA distribution compared to lighter particles. Both results imply that sufficiently small quantum bodies in free fall are distinguishable by their masses contrary to the weak equivalence principle.

\section*{Acknowledgements}
The authors would like to acknowledge J.J.P. Magadan for his help in solving the dynamics of the TOA operator.

\section*{Appendix A: Derivation of the condition for the spread of the wavepacket}
\label{sec:sigma_cond}

Here we show how we quantify the spread of the wavepacket to be sufficiently small as mentioned in Sec. \ref{sec:toa_expec}. The single-peaked wavepacket is centered at the initial position $q=-q_o$ which is non-localized between $q=-q_o-\delta$ to $q=-q_o+\delta$. This wavepacket is then fired upward with the arrival point being the origin and in order for the the TOA to be real-valued to indicate arrival at the origin, then the farthest point of this wavepacket must also have a real-valued TOA at the origin, that is
$1 >2g(q_o+\delta)/v_0^2$ or 
$\delta < v_o^2/2g-q_o$.
It thus follows that, 
$2\delta = \sigma <  v_o^2/g-2q_o$.

\section*{Appendix B: Coarse graining of the time of arrival operator} To study the dynamics of the TOA-operator $\opr{T}$ and to obtain the corresponding time of arrival distribution, one needs to solve the eigenvalue problem for the time of arrival operator $\opr{T}$. However, solving analytically the eigenvalue problem is intractable. The eigenvalue problem is then solved numerically by coarse graining. This is done by confining the system in a large box of length $2l$ centered at the arrival point. The coarse grained version of $\opr{T}$ is then obtained by projecting $\opr{T}$ in the Hilbert space $\mathcal{H}_l=L^2[-l,l]$. The projection of $\opr{T}$ is the integral operator $(\opr{T}_l\varphi)(q)=\int_{-l}^l \left<q|\opr{T}|q'\right>\varphi(q')\,\mathrm{d}q'$, where $\left<q|\opr{T}|q'\right>$ is the full kernel of the TOA-operator $\opr{T}$ in the entire real line. The eigenvalue problem for  $\opr{T}_l$ is then solved by quadrature using Nystrom method. The evolution of the eigenfunctions of $\opr{T}_l$ is detailed in  \cite{jaykel} and \cite{IJMPA}. On the other hand, the construction of the time of arrival distribution is described in  \cite{G2009PRSA} and \cite{toa1}.

\end{document}